\documentstyle[11pt,aaspp4]{article}

\def\farcs{\hbox{$.\!\!^{\prime\prime}$}}
\def\etal {\emph{et~al.}}

\def\go {\mathrel{\raise.3ex\hbox{$>$}\mkern-14mu\lower0.6ex\hbox{$\sim$}}}
\def\lo {\mathrel{\raise.3ex\hbox{$<$}\mkern-14mu\lower0.6ex\hbox{$\sim$}}}
\def\fd {\dot f}
\def\fdd {\ddot f}
\def\fddd {\stackrel {\ldots}{f}}
\def\fdddd {f^{(4)}}
\def\fddddd {f^{(5)}}

\begin{document}

\title{Theoretical Implications of the PSR~B1620$-$26 Triple System and its Planet}

\author{Eric B.~Ford\altaffilmark{1}, Kriten J.~Joshi\altaffilmark{2}, 
 Frederic A.~Rasio\altaffilmark{3,4} and Boris Zbarsky\altaffilmark{5}}

\affil{Department of Physics, Massachusetts Institute of Technology}

\altaffiltext{1}{6-218M MIT, 77 Massachusetts Ave, Cambridge, MA 02139; email: eford@mit.edu.}
\altaffiltext{2}{6-218M MIT, 77 Massachusetts Ave, Cambridge, MA 02139; email: kjoshi@mit.edu.}
\altaffiltext{3}{6-201 MIT, 77 Massachusetts Ave, Cambridge, MA 02139; email: rasio@mit.edu.}
\altaffiltext{4}{Alfred P.\ Sloan Research Fellow.}
\altaffiltext{5}{6-218M MIT, 77 Massachusetts Ave, Cambridge, MA 02139; email: bzbarsky@mit.edu.}

\begin{abstract}

We present a new theoretical analysis of the PSR~B1620$-$26 triple system
in the globular cluster M4, based on the latest radio pulsar timing data, 
which now include measurements of 
five time derivatives of the pulse frequency. These data allow us to determine 
the mass and orbital parameters of the second companion completely (up to the 
usual unknown orbital inclination angle $i_2$). 
The current best-fit parameters correspond to a second companion of planetary mass,
$m_2 \sin i_2 \simeq 7\times10^{-3}\,\rm M_\odot$, in an orbit of eccentricity
$e_2\simeq 0.45$ and semimajor axis $a_2\simeq 60\,\rm AU$. 
Using numerical scattering experiments, we study a possible formation scenario 
for the triple system, which involves 
a dynamical exchange interaction between the binary pulsar and a primordial star--planet 
system. The current orbital parameters of the triple are consistent with 
such a dynamical origin, and suggest that the separation of the parent star--planet
system was very large, $\go 50\,$AU. We also examine the possible origin of the anomalously
high eccentricity of the inner binary pulsar. While this eccentricity could have been 
induced during the same dynamical interaction that created the triple, 
we find that it could equally well
arise from long-term secular perturbation effects in the triple,
combining the general relativistic precession of
the inner orbit with the Newtonian gravitational perturbation of the planet.
The detection of a planet in this system may be taken as evidence that
large numbers of extrasolar planetary systems, not unlike those discovered 
recently in the solar neighborhood, also exist in old star clusters.

\end{abstract}


\section{Introduction}

PSR~B1620$-$26 is a unique millisecond radio pulsar. The pulsar is a member of
a hierarchical triple system located in or near the
core of the globular cluster M4. It is the only radio pulsar known in a triple
system, and the only triple system known in any globular cluster. The inner binary 
of the triple contains the $\simeq 1.4\,M_{\odot}$ neutron star with 
a $\simeq 0.3 M_{\odot}$ white-dwarf
companion in a 191-day orbit (Lyne \etal\ 1988; McKenna \& Lyne 1988)
The triple nature of the system was
first proposed by Backer (1993) in order to explain the unusually high residual 
second and third pulse frequency derivatives left over after subtracting a standard 
Keplerian model for the pulsar binary.

The pulsar has now been timed for eleven years since its discovery
(Thorsett, Arzoumanian, \& Taylor 1993; Backer, Foster, \& Sallmen 1993;
Backer \& Thorsett 1995; Arzoumanian \etal\ 1996; Thorsett \etal~1999). 
These observations have not only confirmed the triple nature of the system, 
but they have also provided tight constraints on the mass and orbital 
parameters of the second companion. Earlier calculations using three pulse
frequency derivatives suggested that the mass of the second companion could
be anywhere between $10^{-3}$ and $1\,M_\odot$, with corresponding
orbital periods in the range $\sim 10^2-10^3\,$yr (Michel 1994; Rasio 1994;  
Sigurdsson 1995). More recent calculations using four frequency derivatives
and preliminary measurements of the orbital perturbations of the inner
binary have further constrained the mass of the second companion,
and strongly suggest that it is a giant planet or a brown dwarf of mass 
$\sim 0.01\,M_\odot$ at a distance $\sim 50\,$ AU from the pulsar 
binary (Arzoumanian \etal\ 1996; Joshi \& Rasio 1997).
In this paper we present a new analysis of the pulsar timing data, 
including the most recent
observations of Thorsett \etal\ (1999; hereafter TACL). The data now include 
measurements of
five pulse frequency derivatives, as well as improved measurements and 
constraints on various orbital perturbation effects in the triple. 

Previous optical observations by Bailyn \etal\ (1994) and 
Shearer \etal\ (1996) using ground-based
images of M4 had identified a possible optical counterpart for the pulsar,
consistent with a $\sim0.5\,M_\odot$ main-sequence star, thus
contradicting the theoretical results, which suggest a much lower-mass
companion. However, it also seemed possible that the object could be a blend
of unresolved fainter stars, if not a chance superpositon. 
Later HST WFPC2 observations of the same region by Bailyn \etal\ (1999) 
have resolved the uncertainty. The much higher resolution ($\sim0.1\farcs$) HST image
shows no optical counterpart at the pulsar position, down to a magnitude 
of ${\rm V}\simeq23$,
therefore eliminating the presence of any main-sequence star in the system. Thus, 
the optical observations are now consistent 
with the theoretical modeling of the pulsar timing data.

PSR~B1620$-$26 is not the first millisecond pulsar system in which a 
planet (or brown dwarf) has been detected. The first one, PSR~B1257+12, is an
isolated millisecond pulsar with three clearly detected inner planets 
(all within 1 AU) of terrestrial masses in circular orbits around the
neutron star (Wolszczan 1994). Preliminary evidence for
at least one giant planet orbiting at a much larger distance from the
neutron star has also been reported (Wolszczan
1996; see the theoretical analysis in Joshi \& Rasio 1997). 
In PSR~B1257+12 it is likely that the
planets were formed in orbit around the neutron star, perhaps out of
a disk of debris left behind following the complete evaporation of the
pulsar's stellar binary companion (see, e.g., Podsiadlowski 1995). 
Such an evaporation process has been observed in several eclipsing
binary millisecond pulsars, such as PSR~B1957+20 (Arzoumanian \etal~1994) and
PSR~J2051-0827 (Stappers \etal~1996), where the companion masses have 
been reduced to $\sim0.01-0.1\,M_\odot$ by ablation. These companions
used to be ordinary white dwarfs and, although their masses are now
quite low, they cannot be properly called either planets or brown dwarfs.

In PSR~B1620$-$26, the hierarchical triple configuration of the system and its
location near the core of a dense globular cluster suggest that the second
companion was acquired by the pulsar following a dynamical interaction with another
object (Rasio, McMillan \& Hut 1995; Sigurdsson 1995). This object could have been 
a primordial binary with a low-mass brown-dwarf component, or a main-sequence star
with a planetary system containing at least one massive giant planet.
Indeed the possibility of detecting ``scavenged'' planets around millisecond
pulsars in globular clusters was discussed by Sigurdsson (1992) even before
the triple nature of PSR~B1620$-$26 was discovered.

Objects with masses $\sim0.001-0.01\,M_\odot$ have recently been detected
around many nearby solar-like stars in Doppler searches for extrasolar planets
(see Marcy \& Butler 1998 for a review).
In at least one case, Upsilon Andromedae, several objects have been detected
in the same system, clearly establishing that they are members of a planetary
system rather than a very low-mass stellar (brown dwarf) binary companion
(Butler \etal~1999). For the second companion of PSR~B1620$-$26,
of mass $\sim0.01\,M_\odot$, current observations and theoretical modeling do
not make it possible to determine whether the object was originally formed as
part of a planetary system, or as a brown dwarf.
In this paper, we will simply follow our prejudice, and henceforth will refer to
the object as ``the planet.''

One aspect of the system that remains unexplained, and can perhaps provide
constraints on its formation and dynamical evolution, is the unusually high 
eccentricity $e_1 = 0.0253$ of the inner binary. This is much larger than 
one would expect for a binary millisecond pulsar formed through the standard
process of pulsar recycling through accretion from a red giant companion. During
the mass accretion phase, tidal circularization of the orbit through turbulent
viscous dissipation in the red-giant envelope should have brought the eccentricity
down to $\lo10^{-4}$ (Phinney 1992). At the same time, however, the measured
value may appear too small for a dynamically induced eccentricity.
Indeed, for an initially circular binary, the eccentricity induced by 
a dynamical interaction with another star is an extremely steep function of the 
distance of closest approach (Rasio \& Heggie 1995). Therefore a ``typical'' interaction
would be expected either to leave the eccentricity unchanged, 
or to increase it to a value of
order unity (including the possibility of disrupting the binary). In addition,
one expects that most ``exchange'' interactions with a star--planet system 
will lead to the ejection of the planet while the star is retained in a bound
orbit around the pulsar binary (Rasio \etal~1995; Heggie, Hut, \& McMillan 1996).
In this paper we will present the results of new numerical scattering experiments
simulating encounters between the binary pulsar and a star--planet system.
These simulations allow us to estimate quantitatively the probability of retaining
the planet in the triple, while perhaps at the same time inducing a small but significant 
eccentricity in the pulsar binary.

Secular perturbations in the triple system can also lead to an increase in the
eccentricity of the inner binary. A previous analysis assuming nearly coplanar orbits
suggested that, starting from a circular inner orbit, an eccentricity as large as 
0.025 could only be induced by the
perturbation from a stellar-mass second companion (Rasio 1994), which is now ruled
out. For large relative
inclinations, however, it is known that the eccentricity perturbations can in
principle be considerably larger (Kozai 1962; see Ford, Kozinsky \& Rasio 1999 
for a recent treatment). In addition, the Newtonian secular perturbations due to the tidal
field of the second companion can combine
nonlinearly with other perturbation effects, such as the general relativistic
precession of the inner orbit, to produce enhanced eccentricity perturbations
(Ford \etal~1999).
In this paper, we re-examine the effects of secular perturbations on eccentricity
in the PSR~B1620$-$26 triple,
taking into account the possibly large relative inclination of the orbits, as well
as the interaction between Newtonian perturbations and the general relativistic
precession of the pulsar binary. 

\section{Analysis of the Pulsar Timing Data}

\subsection{Pulse Frequency Derivatives}

The standard method of fitting a Keplerian orbit to timing residuals
cannot be used when the pulsar timing data cover only a small fraction of
the orbital period (but see TACL for an attempt at fitting
two Keplerian orbits to the PSR~B1620$-$26 timing data). For PSR~B1620$-$26, 
the duration of the observations,
about 11 years, is very short compared to the likely orbital period of
the second companion, $\go100\,$yr. In this case, 
pulse frequency derivatives (coefficients in a Taylor expansion of 
the pulse frequency around a reference epoch) can be derived to characterize the 
shape of the timing residuals (after subtraction of a Keplerian model for the inner
binary). It is easy to show that from {\em five\/} well-measured
and dynamically-induced frequency derivatives one can obtain a complete
solution for the orbital parameters and mass of the companion, up to the
usual inclination factor (Joshi \& Rasio 1997; hereafter JR97).

In our previous analysis of the PSR~B1620$-$26 system (JR97), we used the first 
four time derivatives of the pulse frequency to solve for a one-parameter family
of solutions for the orbital parameters and mass of the second companion.
The detection of the fourth derivative, 
which was marginal at the time, has now been confirmed (TACL). 
In addition, we now also have a preliminary
measurement of the \emph{fifth\/} pulse derivative. This allows us in principle
to obtain a 
unique solution, but the measurement uncertainty on the fifth derivative 
is very large, giving us correspondingly large uncertainties on the
theoretically derived parameters of the system. Equations and details on the method
of solution were presented in JR97, and will not be repeated here.

Our new solution is based on the latest available values of 
the pulse frequency derivatives, obtained by TACL 
for the epoch MJD 2448725.5:
\begin{eqnarray}
{\rm Spin\, Period}\, P & = & 11.0757509142025(18) \,{\rm ms} \nonumber \\
{\rm Spin\, frequency}\, f & = & 90.287332005426(14) \, {\rm s}^{-1} \nonumber \\
\fd & = & -5.4693(3)\times10^{-15}\,  {\rm s}^{-2} \nonumber \\
\fdd & = & 1.9283(14)\times10^{-23} \, {\rm s}^{-3} \\
\fddd & = & 6.39(25)\times10^{-33} \, {\rm s}^{-4} \nonumber \\
\fdddd & = & -2.1(2)\times10^{-40} \, {\rm s}^{-5} \nonumber \\
\fddddd & = & 3(3)\times10^{-49} \, {\rm s}^{-5} \nonumber
\end{eqnarray}
Here the number in parenthesis is a conservative estimate of the formal $1\sigma$ error 
on the measured best-fit value, taking into account the correlations between
parameters (see TACL for details). It should be noted that the best-fit value 
for the fourth derivative quoted earlier by Arzoumanian \etal\
(1996) and used in JR97, $\fdddd = -2.1(6)\times10^{-40} \, {\rm s}^{-5}$,
has not changed, while the estimated $1\sigma$ error has decreased by a factor of three.
This gives us confidence that the new measurement of $\fddddd$, although 
preliminary, will not change significantly over the next few years as more
timing data become available. 

Since the orbital period of the second companion is much
longer than that of the inner binary, we treat the inner binary as a single object. 
Keeping the same notation as in JR97, we let $m_1 = m_{\rm NS} + m_c$ be the mass of the 
inner binary pulsar, with $m_{\rm NS}$ the mass of the neutron star and $m_c$ the mass 
of the (inner) companion, and we denote by $m_2$ the mass of the second companion.
The orbital parameters are the longitude $\lambda_2$ at epoch (measured from 
pericenter), the longitude of pericenter $\omega_2$ (measured from the ascending 
node), the eccentricity $e_2$, semimajor axis $a_2$, and inclination $i_2$ (such 
that $\sin i_2=1$ for an orbit seen edge-on). They all
refer to the orbit of the second companion with respect to the center of 
mass of the system (the entire triple). A subscript 1 for the 
orbital elements refers to the orbit of the inner binary.
We assume that $m_{\rm NS}=1.35\,M_\odot$, giving $m_c\sin i_1 =0.3\,M_\odot$, 
where $i_1$ is the inclination of the pulsar binary (Thorsett \etal\ 1993),
and we take $\sin i_1=1$ for the analysis presented in this section since 
our results depend only very weakly on the inner companion mass.

The observed value of $\fd$ is in general determined by a combination of the
intrinsic spin-down of the pulsar and the acceleration due to the second 
companion. However, in this case, the observed value of $\fd$ has changed
from $-8.1\times10^{-15}\rm{s^{-2}}$ to $-5.4\times10^{-15}\rm{s^{-2}}$
over eleven years (TACL). Since the intrinsic spin-down rate is essentially
constant,
this large observed rate of change indicates that the observed $\fd$ is 
almost entirely acceleration-induced. Similarly, the observed value of $\fdd$ 
is at least an order of magnitude larger than the estimate of $\fdd$
from intrinsic timing noise, which is usually not measurable
for old millisecond pulsars (see Arzoumanian \etal\ 1994, TACL, and JR97). 
Intrinsic contributions to the higher derivatives
should also be completely negligible for millisecond pulsars. 
Hence, in our analysis, we assume that all observed frequency derivatives 
are dynamically induced, reflecting the presence of the second companion.

Fig.~1 illustrates our new one-parameter family of solutions, or
``standard solution,'' obtained using the updated values of the first 
four pulse frequency derivatives. There are no significant differences
compared to the solution obtained previously in JR97.
The vertical solid line indicates the unique solution obtained by including the 
fifth derivative. It corresponds to a second companion mass
$m_2 \sin i_2 = 7.0\times10^{-3}\,\rm M_\odot$, eccentricity
$e_2=0.45$, and semimajor axis $a_2=57\,\rm AU$. For a total system mass
$m_1+m_2=1.65\,M_\odot$ this gives an outer orbital period $P_2=308\,$yr.

It is extremely reassuring to see that the new 
measurement of $\fddddd$ is consistent with the family of solutions obtained
previously on the basis of the first four derivatives. The implication is that
the signs and magnitudes of these five independently measured quantities are
all consistent with the basic interpretation of the data in terms of a 
second companion orbiting the inner binary pulsar in a Keplerian orbit.

For comparison, the two vertical dashed lines in Fig.~1 indicate the
change in the solution obtained by decreasing the value of $\fddddd$ by a
factor of 1.5 (right), or increasing it by a factor of 1.5 (left). Note that 
lower values of $\fddddd$ give higher values for $m_2$. If we
vary the value of $\fddddd$ within its entire $1\sigma$ error bar,
all solutions are allowed, except for the extremely low-mass solutions
with $m_2 \lo 0.002$. In particular, the present $1\sigma$ error on $\fddddd$ does not 
strictly rule out a hyperbolic orbit ($e_2 > 1$) for $m_2$. 
However, it is still possible to derive a strict upper limit on $m_2$ by considering 
hyperbolic solutions and requiring that the relative velocity at infinity of 
the perturber be less than the escape speed from the cluster core. 
This will be discussed in \S 2.3. A strict {\em lower limit\/} on the mass,
 $m_2\go 2\times10^{-4}\,M_\odot$, can be set by requiring that the orbital period of the 
second companion be longer than the duration of the timing observations (about
$10\,$yr). Note that
all solutions then give dynamically stable triples, even at the low-mass,
short period limit (see JR97 for a more detailed discussion).

\subsection{Orbital Perturbations}

Additional constraints and consistency checks on the model can be obtained
by considering the perturbations of the orbital elements of the inner binary
due to the presence of the second companion. These include a precession of
the pericenter, as well as short-term linear drifts in the inclination and eccentricity.
The drift in inclination can be detected through a change in the
projected semimajor axis of the pulsar. The semimajor axis itself is
not expected to be perturbed significantly by a low-mass second companion
(Sigurdsson 1995).

The latest measurements, obtained by adding a linear drift term to each 
orbital element in the Keplerian fit for the inner binary (TACL) give:
\begin{eqnarray}
\dot \omega_1 & = & -5(8)\times10^{-5}\,^{\circ}\,\rm yr^{-1},\\
\dot e_1 & = & 0.2(1.1)\times10^{-15}\,\rm s^{-1}, \\
\dot x_p & = & -6.7(3)\times10^{-13},
\end{eqnarray}
where $x_p = a_p \sin i_1$ is the projected semimajor axis of the pulsar. 
Note that only $\dot x_p$ is clearly detected, while the other two measurements 
only provide upper limits.

We use these measurements to constrain the system by
requiring that all our solutions be consistent with these secular 
perturbations. To do this, we perform Monte-Carlo simulations,
constructing a large number of random realizations of the triple 
system in 3D, and accepting or rejecting them on the basis of
compatibility with the measured orbital perturbations (see JR97
for details). 
The eccentricity of the outer orbit is selected randomly assuming
a thermal distribution, and the other orbital parameters are then
calculated from the standard solutions obtained in \S2.1
(see below for a modified procedure which includes the preliminary
measurement of $\fddddd$). The unknown inclination angles $i_1$ and 
$i_2$ are generated assuming random orientations of the orbital planes, 
and the two position angles of the second companion are determined using
$i_1$, $i_2$, $\omega_2$, $\lambda_2$, and an additional undetermined
angle $\alpha$, which (along with $i_1$ and $i_2$) describes the 
relative orientation of the two orbital planes.
The perturbations are calculated theoretically for each realization 
of the system, assuming a fixed position of the second companion.
The perturbation equations are given in Rasio (1994) and JR97.

Fig.~2 shows the resulting probability distributions for the 
mass $m_2$ and the current separation (at epoch) $r_{12}$
of the second companion.
The Monte-Carlo trials were performed using only
our standard solution based on four pulse frequency derivatives, 
since the fifth derivative
is still only marginally detected. The solid line indicates the
value given by the preliminary measurement of $\fddddd$.
The most probable value of $m_2 \simeq 0.01 M_\odot$ is 
consistent with the range of values obtained from the complete
solution using the fifth derivative. 
The two dashed lines indicate the values obtained from the complete
solution by decreasing the value of $\fddddd$ by a
factor of 1.5 (right), or increasing it by a factor of 1.5 (left).

Fig.~3 shows the probability distribution of $\fddddd$ predicted by our 
Monte-Carlo simulations, with the vertical solid line indicating
the preliminary measurement, and the dashed lines showing 
the values of $\fddddd$ increased and decreased by a factor of 1.5,
as before.
The most probable value is clearly consistent with the measured value,
well  within
the formal $1\sigma$ error. This result provides another independent 
self-consistency check on our model, indicating that all the present timing data,
including the orbital perturbations, are consistent with the basic 
interpretation of the system in terms of a triple.

Since the uncertainty in $\fddddd$ is still large, it can also be 
included in the Monte-Carlo procedure as a new constraint, in addition
to the three orbital perturbation parameters.
For each realization of the system, we now also calculate the predicted value 
of $\fddddd$, and we add compatibility with the measured value (assuming
a Gaussian distribution around the best-fit value, with the quoted $1\sigma$
standard deviation) as a condition to accept the system.
These Monte-Carlo simulations with the additional
constraint on $\fddddd$ produce results very similar to those of Fig.~2, 
giving in particular
the same value for the most probable second companion mass. This is of
course not surprising, given the results of Fig.~3. The only significant 
difference is that the small number of accepted solutions with $m_2 \lo 10^{-3} 
\rm{M_\odot}$ seen in Fig.~2 (small hump seen on the left of the main
peak) are no longer allowed.

We also conducted Monte-Carlo simulations using $\fddddd$ \emph{alone} 
as a constraint, to determine the {\em a priori\/} probability distributions
of the three secular perturbations for solutions that are consistent with
all measured pulse frequency derivatives.
Fig.~4 shows the distribution of the absolute values of the three 
secular perturbations (the symmetry of the problem allows both positive
and negative values with equal probability)
that were consistent with the frequency derivative data.
We see that the allowed values of the
perturbations span several orders of magnitude, with the most probable values
of $\dot x_p$, $\dot e_1$, and $\dot \omega_1$ being roughly consistent with 
their observed values. 

We note from Fig.~4 that
the wide probability distribution for $\dot x_p$ easily
allows (although it does not favor) values much smaller 
than its currently observed value. This suggests the  
possibility that the observed value of $\dot x_p$ may be significantly affected by
external effects, such as the proper motion of the pulsar (which, if significant, can
also lead to a slow drift in the inclination of the orbital plane to the line
of sight, through the change in the direction of the line of sight itself).
Indeed this possibility was discussed by  Arzoumanian \etal\ (1996). They
pointed out that, if the pulsar proper motion is equal to
the published cluster proper motion, $\mu = 19.5\,\rm{mas\,yr^{-1}}$  
(Cudworth \& Hansen 1993)
and if the proper motion alone determines the measured $\dot x_p$,
this would require the inclination angle $i_1$ to be less than 
about $16^{\circ}$, implying a mass $\go 1.4\,M_\odot$ for the (inner)
pulsar companion. While another 
neutron star, or a black hole companion cannot be strictly ruled out,
this would require a highly unusual formation scenario for the system, and
would make the nearly circular orbit of the binary pulsar extremely difficult
to explain. 

Note, however, that the published cluster proper motion may be
in error. Indeed, the pulsar timing proper motion
obtained by TACL,  $\mu = 28.4\,\rm{mas\,yr^{-1}}$, is incompatible with
the published value for the cluster, since it would imply a
relative velocity of the pulsar system with respect to the cluster of 
$78\pm40\,\rm{km}\,{\rm s}^{-1}$ (TACL), 
far greater than the escape speed from the cluster. Since the
location of the pulsar very near the cluster core makes its association with the
cluster practically certain, 
it is very likely that either the timing proper motion or the 
cluster proper motion is incorrect (see TACL for further discussion). 

If we assumed that the observed $\dot x_p$ is due entirely to proper motion, we can
obtain an estimate of the proper motion.
For example, take the inner pulsar companion to be a white dwarf of mass $0.35\,M_\odot$,
corresponding to an inclination $i_1=55^{\circ}$. We find that
this would require a proper motion $\mu \simeq 100\,\rm{mas\,yr^{-1}}$, 
or about five times greater than
the current value of the optical (cluster) proper motion, and implying a velocity
far greater than the escape speed for the cluster. 
But this is also about four times greater than the pulsar timing proper motion
obtained by TACL. Therefore, if the observed $\dot x_p$ were due to proper motion,  
then {\em both\/} the optical and timing proper motion would have to be
incorrect, which seems unlikely. 

\subsection{Hyperbolic Solutions}

Given the uncertainty in the proper motion of the system and the
fact that the measurements of $\fddddd$ and the orbital perturbations
would allow it, we have also considered the possibility that
the ``second companion'' may in fact be an object
on a hyperbolic orbit, caught in the middle of a close interaction with 
the binary pulsar. The probability of observing
the system during such a transient event is of course very low
($\sim 2\times10^{-5}$ if the binary pulsar is in the core of M4;
see Phinney 1993), but it cannot be ruled out a priori.

In Fig.~5, we show the predicted value of $\fddddd$ as a function of the mass of
the second companion ($m_2 \sin i_2$), obtained by extending the solutions of
Fig.~1 into the hyperbolic domain. Values of $m_2\sin i_2 > 0.012\,M_\odot$ 
correspond to a
hyperbolic orbit for the second companion ($e_2 > 1$). 
The solid horizontal line indicates the measured value of $\fddddd$, and the dashed
lines indicate the current $1\,\sigma$ error bar. We see that the current
uncertainty in $\fddddd$ allows the entire range of hyperbolic orbits for the
second companion. 

However, a strict upper limit on $m_2$ can be derived from
the requirement that the pulsar remain bound to the cluster. 
For $m_2 \sin i_2 > 0.055\,M_\odot$, we find that the 
relative velocity (``at infinity'') would exceed the escape speed from the 
cluster core ($\simeq 12\,\rm{km}\,{\rm s}^{-1}$; see Peterson \etal~1995).
To reach a stellar mass, say $m_2 = 0.5\,M_\odot$, would require a relative velocity 
at infinity $\simeq 45\,\rm{km}\,{\rm s}^{-1}$, which would be well above the escape
speed from the cluster. Thus, even in the hyperbolic regime, the mass
of the second companion is constrained to remain substellar.
In addition, we find that the hyperbolic solutions require the present position of
the second companion to be very close to pericenter ($|\lambda_2| < 10^{\circ}$).
This is not surprising, since the hyperbolic solutions are a smooth extension of
the standard bound solutions shown in Fig.~1 (in which $\lambda_2$ is already 
small for solutions with $e_2 \simeq 1$), but it makes the hyperbolic solutions
even more unlikely.

\section{Triple Formation Process}

\subsection{Recycling in the Triple}

One plausible formation scenario for PSR~B1620$-$26 begins with an old neutron star in a
binary system, which has a dynamical interaction with a star--planet system
(Sigurdsson 1993, 1995). The original companion of the neutron star is ejected in 
the interaction, while
the main-sequence star and its orbiting giant planet are retained in orbit around
the neutron star. The planet is typically retained in a wider orbit around the
new inner binary system. The main-sequence star later evolves to become a red giant,
transferring mass onto the neutron star and recycling it into a millisecond 
pulsar, while at the same time the inner orbit is circularized through tidal
dissipation. Thus, in this scenario, the triple system is formed before the
recycling of the neutron star takes place.
 
Sigurdsson (1993) finds through numerical scattering experiments, that in 
approximately $15\%$ of such exchange interactions (in which the main-sequence
star remains bound to the neutron star), the planet is left 
in a bound, wide orbit around the system. In a majority of such cases,
the planet has a final semimajor axis $10-100$ times larger than the initial semimajor 
axis of the neutron star binary, and an eccentricity in the range $0.3-0.7$. 
Sigurdsson (1993) also
finds that the probability of retaining a planet is proportionately higher
if the main-sequence star has more than one planet orbiting it. 
Thus it seems reasonable to consider such an exchange interaction as a possible
formation scenario for PSR~B1620$-$26. 

However, there are two main difficulties with this scenario.
First, it requires that the age of the millisecond pulsar be at most comparable
to  the age of the triple system. Our modeling of the timing data (\S2) indicates
that the semimajor axis of the planet's orbit is $\go50\,$AU.
This makes the outer orbit of the triple ``soft'' in the cluster, implying that
any close interaction with a passing star will disrupt the triple (ejecting
the planet). Since gravitational focusing is negligible for such a large object,
the lifetime of the triple can then be written
\begin{equation}
\tau_d \simeq 8.5 \times 10^7\, {\rm yr}\, \eta_d^2
\left(\frac{<m>}{0.8\,M_\odot} \right) 
\left(\frac{10^4\,M_\odot\,\rm{pc}^{-3}}{\rho} \right)
\left(\frac{5\,\rm{km\,s^{-1}}}{\sigma} \right) 
\left(\frac{50\,\rm{AU}}{a_2} \right)^2. 
\end{equation}
Here $<m>$ is the average stellar mass in the cluster, $\rho$ is the stellar 
density, $\sigma$ is the 3D velocity dispersion, 
$a_2$ is the semimajor axis of the outer orbit, and $\eta_d\equiv a_2/r_d$,
where $r_d$ is the average distance of closest approach for an encounter that will
disrupt the outer orbit. Based on an extensive set of scattering experiments
(see \S 3.2 below), we find that $r_d/a_2\simeq1.4$ and therefore
the factor $\eta_d^2\simeq0.5$. Hence, in the core of M4, 
the expected lifetime of the triple is $\tau_d \sim3\times10^{7}\,$yr. 
This is much shorter than the estimated age of the binary pulsar, 
$\tau_p \go 10^9\,$yr (Thorsett \etal\ 1993).  However, outside the core, the
expected lifetime of the triple can be up to two orders of magnitude 
longer due to the much lower number density of stars in the halo.
For example, at the half-mass radius of M4, the stellar density is down to
$\sim10^2\,M_\odot\,\rm{pc}^{-3}$  and the lifetime becomes $\tau_{\rm hm} \sim10^9\,$yr.
Hence, this scenario would require that the triple system, currently observed
(in projection) near the edge of the cluster core, is in fact on an orbit that 
extends far outside the core,  allowing it to spend most of its lifetime in the 
less dense cluster halo. 

Indeed, Sigurdsson (1995) showed through numerical simulations, that a
triple system with a low-mass second companion, such as PSR~B1620$-$26, 
is more likely to be observed outside the cluster core than inside the core.
Sigurdsson also finds that such triples can easily survive for up to
$2\times10^9\,\rm{yr}$ by spending most of their time outside the core,
but that once they enter the core, they are quickly disrupted. In contrast, triples
with stellar-mass second companions sink to the core much more quickly due to 
dynamical friction (on a timescale $\lo10^9\,\rm{yr}$), and have a 
lifetime of order $10^8\,\rm{yr}$ in the core (Sigurdsson 1995, Rasio \etal~1995).
Thus the fact that PSR~B1620$-$26 is observed just outside the core suggests
that it probably contains a low-mass second companion (which is consistent with 
our timing solution), or that it is a very young system formed less than 
$5\times10^8\,\rm{yr}$ ago (which is difficult to reconcile with the estimated
age of the binary pulsar in this scenario). 

The other difficulty concerns the eccentricity of the pulsar binary. Since the 
circularization of the inner binary takes place after the triple is formed, this
scenario leaves the higher observed eccentricity of the inner binary unexplained. 
To produce the higher eccentricity, a second dynamical 
interaction of the triple with a passing star can be invoked (Sigurdsson 1995).
However, a very close interaction is needed to induce a significant eccentricity 
in an initially circular binary (Rasio, McMillan, \& Hut 1995; see Heggie \& Rasio 
1996 for a general treatment of this problem).  Therefore, such an encounter 
is likely to disrupt the outer orbit in the triple system. An
encounter with a distance of closest approach of $\simeq 2.5\,$AU to the inner 
binary could induce the observed eccentricity. But this is much smaller than the
semimajor axis of the outer orbit, and roughly half of such encounters
will disrupt the outer orbit (Sigurdsson 1995). Such a close interaction 
occurs on average once in $\sim 4\times10^{8}$\,yr.
Therefore, for each interaction that could have produced the eccentricity of the 
inner binary, we expect $\sim 10$ encounters that could have disrupted the outer 
orbit, leaving the probability of surviving at $\lo 0.01$. 

Alternatively, it is possible that the eccentricity of the inner orbit 
could have been induced later through secular perturbations in the triple. 
Rasio (1994) argued that this is unlikely for a low-mass second companion
with low relative inclination between the two orbits.
However, we find that under certain conditions,
the inner eccentricity can be explained as arising from the combined effects of 
the tidal perturbation by the second companion and the general relativistic 
precession of the inner orbit. We discuss this in more detail in \S\ 4.

\subsection{Pre-Existing Binary Pulsar}

We now propose an alternative formation scenario, which
involves a dynamical exchange interaction between a {\em pre-existing\/} binary 
millisecond pulsar and another wider system containing a giant planet in orbit 
around a main-sequence star. This interaction could lead to the ejection of the 
main-sequence star, leaving the planet in a wide orbit around the binary
pulsar. An exchange interaction of this type might also simultaneously
induce the observed eccentricity of the binary pulsar if either the main-sequence
star or the planet passes sufficiently close to the binary pulsar during the
interaction.
This scenario is a natural extension of the mechanism studied by
Rasio \etal~(1995) for the formation of a stellar triple containing a
millisecond pulsar. Here, however, an added difficulty is that one
expects the planet, and not the main-sequence star, to be preferentially
ejected during the interaction.

To study this binary-binary interaction scenario quantitatively, we have 
performed numerical scattering experiments 
similar to those done by Rasio, McMillan, \& Hut (1995) for stellar triples. 
The binary 
pulsar was treated as a single body of mass $1.65\,$M$_\odot$, scattering 
off several different binary systems containing a  
main-sequence star with a very low-mass companion (which we refer to 
henceforth as the ``star--planet system'').  
The assumption that one can treat the inner pulsar
binary as a single mass is justified because its semimajor axis
($0.77\,$AU) is considerably  smaller than the orbital radius of the planet
in all relevant cases.  In addition, close approaches to the binary pulsar
by either the planet or the main-sequence star are forbidden, since they
would induce an eccentricity larger than the one observed.
For example, an
approach by a $0.01\,M_\odot$ object to within about 1.2$\,$AU (1.5
semimajor axes) of the pulsar binary would be enough to induce an
eccentricity greater than $0.025$ (cf.\ Heggie \& Rasio 1996). Hence in all relevant 
interactions, the pulsar binary could be treated as a single point mass.

Four different outcomes are possible for each simulated interaction:
(1) ionization; (2) no exchange (as in a flyby); (3) exchange in which the 
planet is retained around the binary pulsar; (4) exchange in which the star 
remains bound to the binary pulsar. Exchanges can be further 
divided into ``resonant'' and ``nonresonant'' or ``direct.'' 
In a resonant exchange interaction, all stars
involved remain together for a time long compared to the initial
orbital periods (typically $\sim10^2-10^3$ dynamical times), leading
eventually to the ejection of one object while (in this case) the others
remain in a bound hierarchical triple configuration. In direct
exchange interactions, either the planet or the main-sequence star is 
ejected promptly, following a close approach by the binary pulsar.
In general, the type of outcome depends on the impact parameter, the 
mass of the planet, and the initial planet-star separation.

Heggie, Hut, \& McMillan (1996) have computed cross sections for exchange
interactions in binary-single-star encounters, both numerically using
scattering experiments, and analytically in various limiting regimes.
They include results for binary mass ratios as extreme as $\sim 0.01$.
However, their results are all obtained for {\em hard\/} binaries,
whereas the star--planet system in our scenario represents a very soft
binary. Therefore, we do not expect the results of Heggie \etal~(1996)
to be directly applicable here, although they do provide some qualitative
predictions. For example, for a binary mass ratio $m_2/m_1\simeq0.01$, and 
$m_3/m_1=1$, the ratio of cross sections for ejection of the star ($m_1$)
to ejection of the planet ($m_2$) is about 0.12 for direct exchanges and
0.04 for resonant exchanges, with the total cross section for all resonant
exchanges being about 2.5 times larger than that of all direct exchanges.
For soft binaries we find that direct exchanges are dominant  
and that the relative probability of ejecting the more massive binary member
is higher (see below).

Our scattering experiments were performed using the {\bf scatter3}
module of the STARLAB stellar dynamics package (McMillan \& Hut 1996; 
Portegies Zwart \etal\ 1997).
The mass of the main-sequence star was held constant at $0.8\,M_\odot$,
close to the main-sequence turn-off point of the cluster. The planet was placed on an 
initially circular orbit. The relative velocity at infinity for all encounters 
was taken to be $4\,{\rm km}\,{\rm s}^{-1}$, slightly lower than the 3D velocity 
dispersion in the cluster core ($\simeq 6\,{\rm km}\,{\rm s}^{-1}$; see 
Peterson \etal~1995).
The impact parameter $b$ was randomly selected between zero
and $b_{\rm max}$, with probability $\rm{p}(b) \propto 2\pi b$.
The upper limit was selected such that all interactions with 
$b \ge b_{\rm max}$ led to a fly-by with no exchange. In units of the
initial semimajor axis of the star--planet system, $b_{\rm max}$ varied
from about 5 to 35 depending on the mass of the planet.
Three different masses were selected for the ``planet:'' 0.001 M$_\odot$, 
0.01 M$_\odot$, and 0.1 M$_\odot$. In each case, the semimajor 
axis of the star--planet system
was initially chosen to be $5\,$AU, since this is the typical distance at 
which giant planets are expected to form. In addition, in an effort to
better match the orbital parameters of triple systems like PSR~B1620$-$26, 
we also considered initial star--planet semimajor axes of $30\,$AU and 
$50\,$AU. 

We first address the possibility of capturing the planet into a wide
orbit around the pulsar binary, while ejecting the star during an
exchange interaction. Fig.~6 shows the distributions of the
semimajor axis and eccentricity of the planet in the triple for 
those interactions in which the planet was captured and the binary 
pulsar was not disrupted. We see that the final semimajor axis of the
planet in the triple is likely to be within a factor of $\sim2$ of the
initial star--planet separation $a_p$ (Note that, in this section, an
index p refers to the orbit of the star--planet system, and not to
the pulsar, as in \S 2). The eccentricity distribution is 
very broad, and approaches a ``thermal'' distribution (${\rm p}(e)\propto e$)
for very wide initial separations.

Fig.~7(a)-(e) shows the corresponding probabilities of the 
three possible outcomes (branching ratios) for different
planet masses and initial separations of the star--planet system.
The distance of closest approach $r_p$ is
given in units of the initial star--planet semimajor axis $a_p$.
Since the planet was the lowest-mass object participating in the
interaction, it was expected to have a much higher probability of being 
ejected. However, we find that 
the branching ratio for interactions resulting in the capture of the planet
remains significant for all interactions close enough to
result in an exchange. 
For example, from Fig.~7(e), we see that for an interaction between
the binary pulsar and a star--planet system with separation 50 AU
and planet mass $m_p = 0.01\,\rm{M_\odot}$, the branching ratio for 
planet capture (ejection of the star) is in the range $\simeq10-30\%$ 
for distances of closest approach in the range $\sim 10-50\,$AU.
For this case, there are no resonant interactions, implying that 
retention of the star has negligible probability, while the branching
ratio for ionization is in the range $\simeq20-60\%$.
The remainder of the interaction cross section ($10\%$ to $70\%$) 
corresponds to ``clean'' flyby's (without capture of either 
star or planet, or ionization). 

To focus our study on interactions in which the final triple configuration
is similar to that of PSR~B1620$-$26, we now eliminate cases in which
the eccentricity induced in the binary pulsar's orbit would have been
larger that the present value of 0.025. The induced eccentricity 
is estimated based on the results of Heggie \& Rasio (1996). 
In making these estimates, all unknown angles, including the 
relative inclination of the two orbits, are chosen randomly
(assuming random orientations). 
Frames (f)--(j) in Fig.~7 show the branching
ratios for those interactions in which the eccentricity induced in 
the binary pulsar during the interaction is below 0.025.
The main difference with
the general results (frames (a)--(e) in Fig.~7)
is that the branching ratio for ejection of the star now goes to zero
for very close interactions, since those interactions in which the star
passes very close to the binary pulsar would have induced an eccentricity
larger than allowed. Fig.~8 shows the corresponding distributions
of semimajor axis and eccentricity of the planet in the triple
(like Fig.~6 but with the additional constraint on induced eccentricity).
Here the only significant difference is in the lower probability of 
retaining the planet on a highly eccentric orbit, especially for more massive
planets (compare, e.g., Fig.~8c to Fig.~6c).

In Table~1, we list the cross sections for the various outcomes, obtained
from our scattering experiments for each of the five cases. The cross sections
are given in units of the initial geometrical cross section of the star-planet 
binary ($\pi a_p^2$). 

We now turn to the question of whether the eccentricity 
of the binary pulsar could have been increased significantly during the
interaction. 
Fig.~9 shows the overall distribution of the final eccentricity 
induced in the binary pulsar following an exchange interaction in which the
planet was retained. In every case, we see
that the induced eccentricity varies over a large range, including the 
currently observed eccentricity of the binary pulsar.
Thus, a scenario in which an exchange interaction successfully 
forms the triple {\em and\/} simultaneously induces the inner binary eccentricity 
we see today is certainly possible, although the present value of the 
eccentricity is not liklely in this scenario.  Alternatively,
the triple could have formed with an initial inner binary eccentricity 
{\em lower\/} than the present value. 
The eccentricity could then be raised to the present value by 
long-term secular perturbation effects (\S 4). Forming the triple with
an initial inner binary eccentricity {\em higher\/} than currently observed
and later reducing it through secular perturbations is also possible.

Our analysis of the PSR~B1620$-$26 timing data (\S 2) indicate that the second 
companion is in a wide orbit around the pulsar binary, with a semimajor 
axis $\go 50\,$AU. 
The results of our scattering experiments indicate that the final semimajor 
axis of the planet in the triple should be comparable to the initial star--planet
separation (see Figs.~6 and~8). 
Hence, the interactions that could have produced a system like 
PSR~B1620$-$26 are those involving a wide 
system with a planet mass $m_p\sim 0.01 M_\odot$ and an 
initial separation $a_p\sim 50$ AU, as shown in
Figs.~7(j) and 8(e). Just like the triple itself, such a wide 
star--planet system is ``soft'' (binding
energy smaller than the typical center-of-mass energy of an encounter) and
has a large interaction cross section, implying a 
short lifetime in the cluster core, $\tau_d \sim 3\times10^{7}\,$yr (cf.\ eq.~5).
Well outside of the cluster core, its lifetime could be much longer,
but since an exchange interaction with another binary is likely to take place
only in the core, it requires the star--planet system to first drift into the core
through dynamical friction. The interaction that destroyed the star--planet system
once it entered the
core could be the one that produced the PSR~B1620$-$26 triple.

\section{Secular Eccentricity Perturbations}

We now examine in detail the possibility that the inner binary eccentricity was 
induced by the secular gravitational perturbation of the second companion. 
In the hierarchical three-body problem, analytic expressions for the maximum
induced eccentricity and the period of long-term eccentricity oscillations are
available in certain regimes, depending on the eccentricities and relative inclination
of the orbits.

\subsection{Planetary Regime}

For orbits with small eccentricities and a small relative inclination
($i\lo 40^{\circ}$), the classical solution for the long-term secular
evolution of eccentricities and longitudes of pericenters can be written
in terms of an eigenvalue and eigenvector formulation
(Brouwer \& Clemence 1961; for an excellent pedagogic summary, see 
Dermott \& Nicholson 1986). This classical solution is valid to all orders
in the ratio of semimajor axes. In this regime the eccentricities oscillate
as angular momentum is transferred between the two orbits.  The
precession of the orbits (libration or circulation)  is coupled to the
eccentricity oscillations, but the relative inclination remains approximately
constant.  In this regime it can be shown that a stellar mass second 
companion would be necessary to induce the observed eccentricity in the inner binary
(Rasio 1994; 1995).  Such a large mass for the third body has been ruled out
by recent timing data (See \S 2), implying that secular perturbations
from a third body in a nearly coplanar and circular orbit does not
explain the observed inner eccentricity.  The lack of a stellar-mass
companion is also supported by the absence of an optical counterpart
for the pulsar in the HST observations by Bailyn \etal~(1999).

Sigurdsson (1995) has suggested that it may be possible for the secular
perturbations to grow further because of random distant interactions of the triple
with other cluster stars at distances $\sim100$\,AU, which would perturb
the long-term phase relation between the inner and outer orbits, allowing
the inner eccentricity to ``random walk'' up to a maximum value up to two orders 
of magnitude 
higher than that predicted for an isolated triple.
However, the current most probable solution from the timing data requires the
the second companion to be in a very wide orbit (separation $\go40$\,AU), giving
it a very short lifetime in the core, and leaving it extremely vulnerable
to disruption by repeated weak encounters (see JR97 for a more detailed discussion).

\subsection{High-Inclination Regime}

For a triple system formed through a dynamical interaction, there is no reason
to assume that the relative inclination of the two orbits should be small.
When the relative inclination of the two orbital planes is $\go
40^{\circ}$, a different regime of secular perturbations is encountered.
This regime has been studied in the past using the quadrupole
approximation (Kozai 1962; see Holman, Touma, \& Tremaine 1987 for a recent
discussion). Here the relative inclination $i$ of the two
orbits and the inner eccentricity $e_1$ are coupled by the integral
of motion (Kozai's integral) 
\begin{equation}
\Theta = (1-e_1^2) \cos^2 i.
\end{equation}
Thus, the
amplitude of the eccentricity oscillations is determined
by the relative inclination. It can be shown that large-amplitude eccentricity
oscillations are possible only when $\Theta < 3/5$ (Holman \etal~1997).
For an initial eccentricity $e_1\simeq0$ and initial
inclination $i_0$ this implies $i_0 > \cos^{-1}\sqrt{3/5}\simeq 40^\circ$
and the maximum eccentricity is then given by
\begin{equation}
e_{1\rm max}= \left(1- \frac{5}{3}\cos^2 i_0 \right)^{1/2}, 
\end{equation}
which approaches
unity for relative inclinations approaching $90^{\circ}$.  
For a sufficiently large relative inclination, this suggests
that it should always be possible to induce an arbitrarily large
eccentricity in the inner binary, and that this could provide an
explanation for the anomalously high eccentricity of the binary
pulsar in the PSR~B1620$-$26 system (Rasio, Ford, \& Kozinsky 1997).
However, there are two additional conditions that must be
satisfied for this explanation to hold.

First, the timescale for reaching a high eccentricity must be
shorter than the lifetime of the system. Although the masses,
initial eccentricities, and ratio of semimajor axes do not affect the maximum
inner eccentricity, they do affect the period of the eccentricity
oscillations (see Holman \etal\ 1997; Mazeh \etal\ 1996).  
The inner longitude of periastron precesses with this period, which can be
quite long, sometimes exceeding the lifetime of the system in the cluster. 
The masses also affect the period, but they decrease the
amplitude of the eccentricity oscillations only when the mass ratio
of the inner binary approaches unity (Ford, Kozinsky, \& Rasio 1999). 
In Fig.~10 we compare the period of the eccentricity oscillations
to the lifetime of the triple in M4 (see \S 4.3 for details on how the
various estimates were obtained). It is clear that for most solutions
the timescale to reach a large eccentricity exceeds the lifetime of the triple.
The only possible exceptions are for very low-mass planets 
($m_2\lo 0.002\,M_\odot$) and with the triple
residing far outside the cluster core. These cannot be excluded, 
but are certainly not favored by the observations (see \S 2).

The second problem is that other sources of perturbations may become significant
over these long timescales. In particular, for an inner binary containing compact
objects, general relativistic effects can become important. 
This turns out to play a crucial role for the PSR~B1620$-$26 system,
and will be discussed in detail in the next section.

\subsection{General Relativistic Effects}

Additional perturbations which alter the longitude of periastron can
indirectly affect the evolution of the eccentricity of the inner
binary in a hierarchical triple.  
For example, tidally- or rotationally-induced quadrupolar 
distortions, as well as general relativity, can cause a significant precession 
of the inner orbit for a sufficiently compact binary.  
If this additional precession is much
slower than the precession due to the secular perturbations, then the
eccentricity oscillations are not significantly affected.  However, if
the additional precession is faster than the secular perturbations,
then eccentricity oscillations are severely damped (Holman \etal\ 
1997; Lin \etal\ 1998).  In addition, if the two precession periods are 
comparable,
then a type of resonance can occur that leads to a significant increase in the
eccentricity perturbation (Ford \etal~1999).  

Fig.~10 compares the various precession periods for PSR~B1620-26 as a
function of the second companion mass $m_2$ for the entire 
one-parameter family of standard solutions constructed in \S 2.
Also shown for comparison is the lifetime of the triple, both in the cluster
core and at the half-mass radius. We assumed that the stellar densities
in the core and at the half-mass radius are 
$10^4\, M_{\odot}\,{\rm pc}^{-3}$ and $10^2\, M_{\odot}\,{\rm pc}^{-3}$, 
with average stellar masses $<m>=0.8\,M_\odot$ and $0.3\,M_\odot$,
respectively, and set the 3D velocity dispersions to 
$6\,{\rm km}\,{\rm s}^{-1}$ in the core and $3\,{\rm km}\,{\rm s}^{-1}$
at the half-mass radius. Disruption of the triple 
is assumed to occur for any encounter with pericenter distance smaller
than $a_2$ (corresponding to setting $\eta_d=1$ in eq.~5). The timescale 
between such encounters was then estimated from simple kinetic theory
(see, e.g., eq.~8-123 of Binney \& Tremaine 1987), taking into
account gravitational focusing (which is significant at the low-mass,
short-period end of the sequence of solutions; instead, our eq.~5 is
valid in the limit where gravitational focusing is negligible).
The precession period $P_{{\rm High}~i}$ for high-inclinations was
calculated using the approximate analytic expression given by 
Holman \etal~(1997, eq.~3), while $P_{{\rm Low}~i}$
is from Brouwer \& Clemence (1961; see Rasio 1995 for simplified 
expressions in various limits). The general relativistic precession
period, $P_{\rm GR}$ is derived, e.g., in Misner, Thorne, \& Wheeler (1973; \S40.5).
Note that, although we have labeled the plot assuming $\sin i_2=1$, only 
$P_{{\rm Low}~i}$ has an explicit dependence on $m_2$ (and it is calculated
here for $\sin i_2=1$).

For nearly all solutions we find that the general relativistic precession
is {\em faster\/} than the precession due to the Newtonian secular perturbations.
The only exceptions are for low-mass second companions ($m_2\lo
0.005\,M_\odot$) in low-inclination orbits. However, we have already
mentioned above (\S 4.1) that in this case the maximum induced eccentricity
cannot reach the present observed value (for detailed calculations, see 
Rasio 1994, 1995, and \S 4.4).

Most remarkably, we see immediately in Fig.~10 that for the most probable 
solution (based on the
current measured value of $\fddddd$ and indicated by the vertical solid line 
in the figure), the two precession periods
are {\em nearly equal\/} for a low-inclination system. This suggests that
resonant effects may play an important role in this system, a
possibility that we have explored in detail using numerical integrations.
We describe the numerical results in the next section.

\subsection{Numerical Integrations of the Secular Evolution Equations}

If the quadrupole approximation used for the high-inclination regime
is extended to octupole order, the resulting secular perturbation equations  
approximate very well the long-term dynamical evolution of hierarchical 
triple systems for a wide range of masses, eccentricities, and inclinations (Ford, 
Kozinsky, \& Rasio 1999).  We have used the octupole-level secular perturbation
equations derived by Ford \etal~(1999) to study the long-term eccentricity
evolution of the PSR~B1620$-$26 triple. We integrate
the equations using the variables $e_1 \sin \omega_1$, $e_1 \cos \omega_1$, 
$e_2 \sin \omega_2$, and $e_2 \cos \omega_2$, where $\omega_1$ and $\omega_2$ 
are the longitudes
of pericenters. This avoids numerical problems for near-circular orbits,
and allows us to incorporate easily the
first-order post-Newtonian correction into our integrator, which is
based on the Burlish-Stoer integrator of Press \etal\ (1992).

We assume that the present inner eccentricity is due entirely to the 
secular perturbations and that the initial eccentricity of the binary pulsar
was much smaller than its present value.
In addition, we restrict our attention to the standard one-parameter
family of solutions constructed in \S 2. From the numerical integrations
we can then determine the
maximum induced eccentricity, which depends only on the relative inclination.  

In Fig.~11 we show this maximum induced eccentricity in the
inner orbit as a function of the second companion mass for several
inclinations.  For most inclinations and masses, we see that the maximum
induced eccentricity is significantly smaller than
that the observed value, as expected from the discussion of \S 4.3.  
However, for a small but significant range of masses near the most
probable value (approximately 
$0.0055\,\rm{M_\odot} < m_2 < 0.0065\,\rm{M_\odot}$),
the induced eccentricity for low-inclination systems can reach values
$\go 0.02$.

If the initial eccentricity of the outer orbit had been larger
than presently observed (as calculated in Fig.~1) but later
decreased through secular perturbations, the evolution of
the inner orbit may have been different.
We have performed numerical integrations for this case as well and
obtained results very similar to those of Fig.~11.  The ``resonance peaks''
are slightly broader to the right side and the maximum induced
eccentricity is somewhat larger when the second companion mass is
large.  In particular, we find that an inner eccentricity $\go 0.02$ 
can then be achieved for masses as high as $m_2 \simeq 0.012\,M_{\odot}$.  
However, from the results of \S 2 we see that
the corresponding orbital separation in the standard solution is extremely large,
$a_2 \go 500\,$AU, and the outer eccentricity $e_2 \go 0.95$, implying a
very short lifetime for the triple in M4, $\tau_d< 10^6\,$yr.

For relative inclinations $50^{\circ} \lo i \lo 70^{\circ}$ we do not
find a peak in the maximum induced eccentricity as a function of $m_2$
consistent with the timing data.  For inclinations $\go 75^{\circ}$ we
again find a peak in the maximum induced eccentricity which
becomes smaller and moves towards larger masses as the relative
inclination of the two orbits is increased.  A significant
induced eccentricity in the inner orbit is also possible for $m_2 \lo 0.004\,M_{\odot}$ 
if the two orbits are very nearly orthogonal (cf.\ \S4.3).  However, the
results of our Monte-Carlo simulations incorporating the
measured value of $f^{(5)}$ do not support this scenario.

As already pointed out in \S 4.3, the maximum induced eccentricity may 
also be limited by the lifetime of the
triple system. From Fig.~10 we see that near resonance we expect 
$P_{{\rm Low}~i}\approx P_{\rm GR}\sim 10^7\,$yr, which is comparable to
the lifetime of the triple in the cluster core, and much shorter than the lifetime
outside of the core. For solutions near a resonance the inner
eccentricity $e_1$ initially grows linearly at approximately the same rate
as it would without the general relativistic perturbation.  However, the
period of the eccentricity oscillation can be many times the period
of the classical eccentricity oscillations.  Although this allows the
eccentricity to grow to a larger value, the timescale for
this growth can then be longer than the expected lifetime of the triple
in the cluster core. For example, with $m_2=0.006\,M_\odot$
we find that the inner binary reaches an eccentricity
of 0.025 after about 1.5 times the expected lifetime of the triple
in the core of M4. Thus, even if the system is near resonance, it
must still be residing somewhat outside the core for the secular
eccentricity perturbation to have enough time to grow to the currently
observed value.

\section{Summary}

Our theoretical analysis of the latest timing data for PSR~B1620$-$26
clearly confirms the triple nature of the system. Indeed, the values of
all five measured pulse frequency derivatives are consistent with our basic 
interpretation of a binary pulsar perturbed by the gravitational influence of a
more distant object on a bound Keplerian orbit.
The results of our Monte-Carlo simulations
based on the four well-measured frequency derivatives 
and preliminary measurements of short-term orbital perturbation effects
in the triple are consistent with the complete solution obtained 
when we include the preliminary measurement of the fifth frequency 
derivative. This complete solution corresponds to a 
second companion of mass
$m_2 \sin i_2 \simeq 7\times10^{-3}\,\rm M_\odot$ in an orbit of eccentricity
$e_2\simeq 0.45$ and semimajor axis $a_2\simeq 60\,\rm AU$ (orbital period
$P_2\simeq300\,$yr). 
Although the present formal $1\sigma$ error on $\fddddd$ is large, we do not
expect this solution to change significantly as more timing data become
available.

At least two formation scenarios are possible for the triple system,
both involving dynamical exchange interactions between binaries in the
core of M4. In the one scenario that we have studied in detail,
a pre-existing binary millisecond pulsar has a dynamical interaction
with a wide star--planet system which leaves the planet bound to the
binary pulsar while the star is ejected. From numerical
scattering experiments we find that the probability of retaining
the planet, although smaller than the probability of retaining the star,
is always significant, with a branching ratio $\simeq10\%-30\%$ for encounters
with pericenter distances $r_p$ in the range $0.2-1a_p$, 
where $a_p\sim50\,$AU is the initial star--planet separation.
All the observed parameters of the triple system are consistent with
such a formation scenario, which also allows the age of the millisecond pulsar
(most likely $\go10^9\,$yr) to be much larger than the lifetime of the triple
(as short as $\sim 10^7\,$yr if it resides in the core of the cluster).

It is also possible that the dynamical interaction that formed the triple
also perturbed the eccentricity of the binary pulsar to the anomalously
large value of $0.025$ observed today. However, we have shown that
through a subtle interaction between the general relativistic corrections
to the binary pulsar's orbit and the Newtonian gravitational perturbation
of the planet, this eccentricity could also have been induced by long-term
secular perturbations in the triple after its formation. The interaction
arises from the near equality between the general relativistic precession
period of the inner orbit and the period of the Newtonian secular
perturbations for a low-inclination system. It allows the eccentricity 
to slowly build up to the presently observed value, on a timescale that
can be comparable to the lifetime of the triple.

All dynamical formation scenarios have to confront the problem that the
lifetimes of both the current triple and its parent star--planet system
are quite short, typically $\sim10^7-10^8\,$yr as they approach
the cluster core, where the interaction is most likely to occur. Therefore,
the detection of a planet in orbit  around the PSR~B1620$-$26 binary
clearly suggests that large numbers of these star--planet systems
must exist in globular clusters, since most of them will be destroyed before
(or soon after) entering the core, and most planets will not be able to
survive long in a wide orbit around any millisecond pulsar system
(where they may become detectable through high-precision pulsar timing).
Although a star--planet separation $a_p\sim50\,$AU may seem quite large
when compared to the orbital radii of all recently detected extrasolar planets
(which are all smaller than a few AU; see Marcy \& Butler 1998), one must
remember that the current Doppler searches are most sensitive to planets in 
short-period orbits, and that they could never have detected a low-mass 
companion with an orbital period $\gg 10\,$yr. In addition, it remains of course
possible that the parent system was a primordial binary star with a low-mass,
brown dwarf component, and not a main-sequence star with planets.

\acknowledgements

We are very grateful to Steve Thorsett for many useful discussions and 
for communicating results of observations in progress. We thank Piet Hut 
and Steve McMillan for providing us with the latest version of the
STARLAB software. 
We are also grateful to C.\ Bailyn and S.\ Sigurdsson for helpful comments,
and to J. Bostick for help with the calculations of \S 3.2. 
This work was supported in part by NSF Grant AST-9618116 and a NASA ATP Grant at MIT. 
F.A.R.\ was supported in part by an Alfred P.\ Sloan Research Fellowship.
E.B.F. was supported in part by the Orloff UROP Fund and the UROP 
program at MIT. F.A.R. also thanks the Theory Division of the Harvard-Smithsonian
Center for Astrophysics for hospitality. 
This work was also supported by the National Computational Science Alliance 
under Grant AST970022N and utilized the SGI/Cray Origin2000 supercomputer
at Boston University.

\clearpage
\begin{deluxetable}{l|lll|lll}
\footnotesize
\tablecaption{Interaction cross sections \label{tbl-1}}
\tablewidth{0pt}
\tablehead{
\colhead{} & \colhead{}   & \colhead{Normal}   & 
\colhead{} & \colhead{}   & \colhead{$e_{\rm induced} < 0.025$}   & 
\colhead{} \nl
\colhead{case} & \colhead{star ejection}   & \colhead{planet ejection}   & 
\colhead{ionization} & \colhead{star ejection}   & \colhead{planet ejection}   & 
\colhead{ionization}
} 
\startdata
(a) 5 AU, 0.001 M$_\odot$  & 1.13$\pm$0.01  	&  0.00$\pm$0.00   	&  1.66$\pm$0.01	& 0.13$\pm$0.07 	& 0.00$\pm$0.00 	& 0.68$\pm$0.03 \nl

(b) 5 AU, 0.01 M$_\odot$   & 1.19$\pm$0.01 	& 0.13$\pm$0.05 	& 1.52$\pm$0.02		& 0.12$\pm$0.07 	& 0.00$\pm$0.00  	& 0.63$\pm$0.03 \nl

(c) 5 AU, 0.1 M$_\odot$    & 1.05$\pm$0.02	& 2.45$\pm$0.01 	& 0.0000$\pm$0.00	& 0.14$\pm$0.07		& 0.79$\pm$0.03 	& 0.00$\pm$0.00  \nl

(d) 30 AU, 0.001 M$_\odot$ & 0.007$\pm$0.004	& 0.000$\pm$0.000 	& 0.008$\pm$0.004	& 0.001$\pm$0.001 	& 0.000$\pm$0.000	& 0.003$\pm$0.001 \nl

(e) 50 AU, 0.01 M$_\odot$  & 0.001$\pm$0.001 	& 0.000$\pm$0.000	& 0.003$\pm$0.002	& 0.0002$\pm$0.0001 	& 0.000$\pm$0.000 	& 0.0007$\pm$0.0002 \nl

\enddata
 
\tablenotetext{}{Cross sections for the various outcomes for the cases described
in \S3.2. In each case, the velocity of the incoming star at infinity was $4\rm{kms^{-1}}$.}
 
\end{deluxetable}

\clearpage
\begin{figure}[t]
\plotone{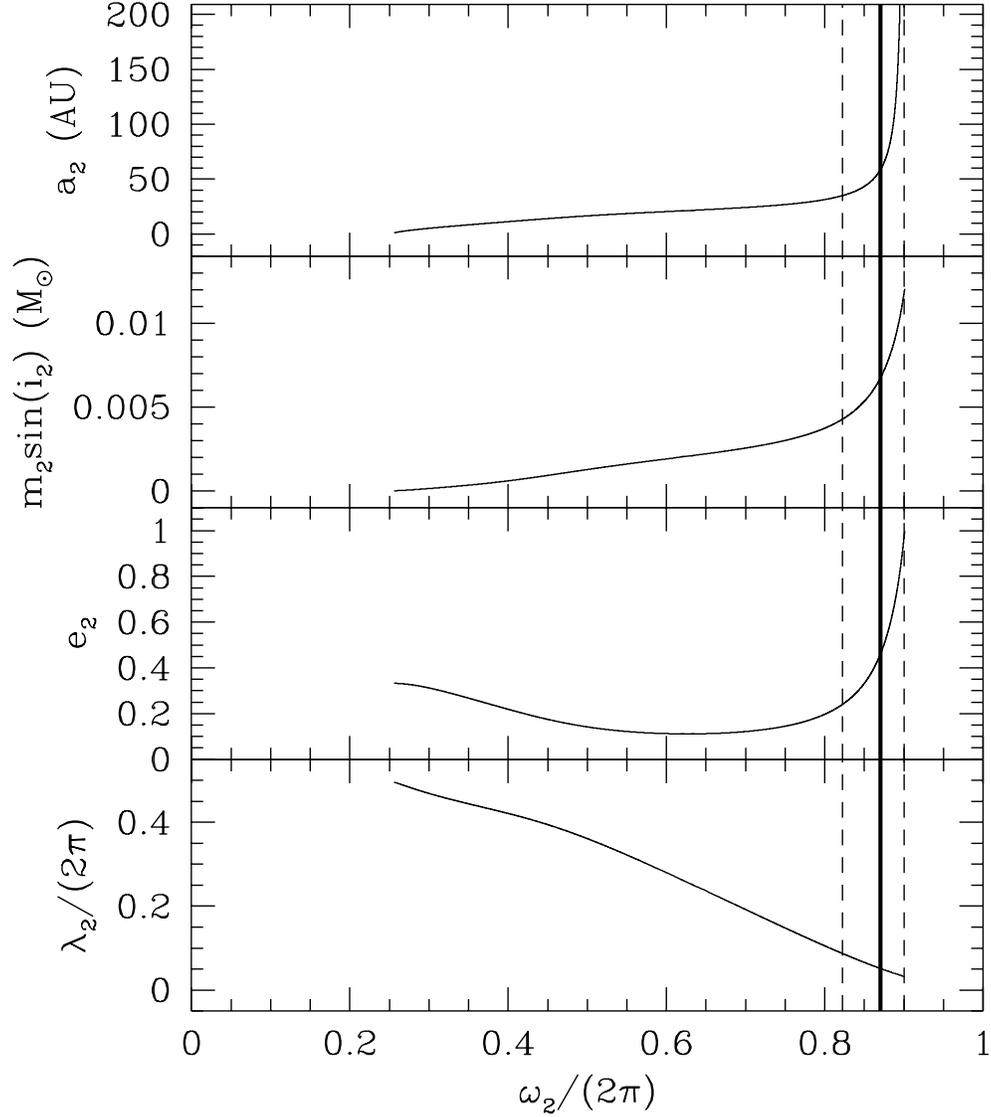}
\caption{Allowed values of the semimajor axis $a_2$, mass $m_2$,
eccentricity $e_2$, longitude at epoch $\lambda_2$, and longitude
of pericenter $\omega_2$
for the second companion of PSR~B1620$-$26, using the latest 
available values for four pulse frequency derivatives. 
The vertical solid line indicates the complete solution obtained by 
including the preliminary value of the fifth derivative. 
The two dashed lines indicate the solutions obtained by decreasing 
the value of $\fddddd$ by a factor of 1.5 (right), or increasing it 
by a factor of 1.5 (left). 
\label{fig1}}
\end{figure}

\clearpage 
\begin{figure}[t]
\plotone{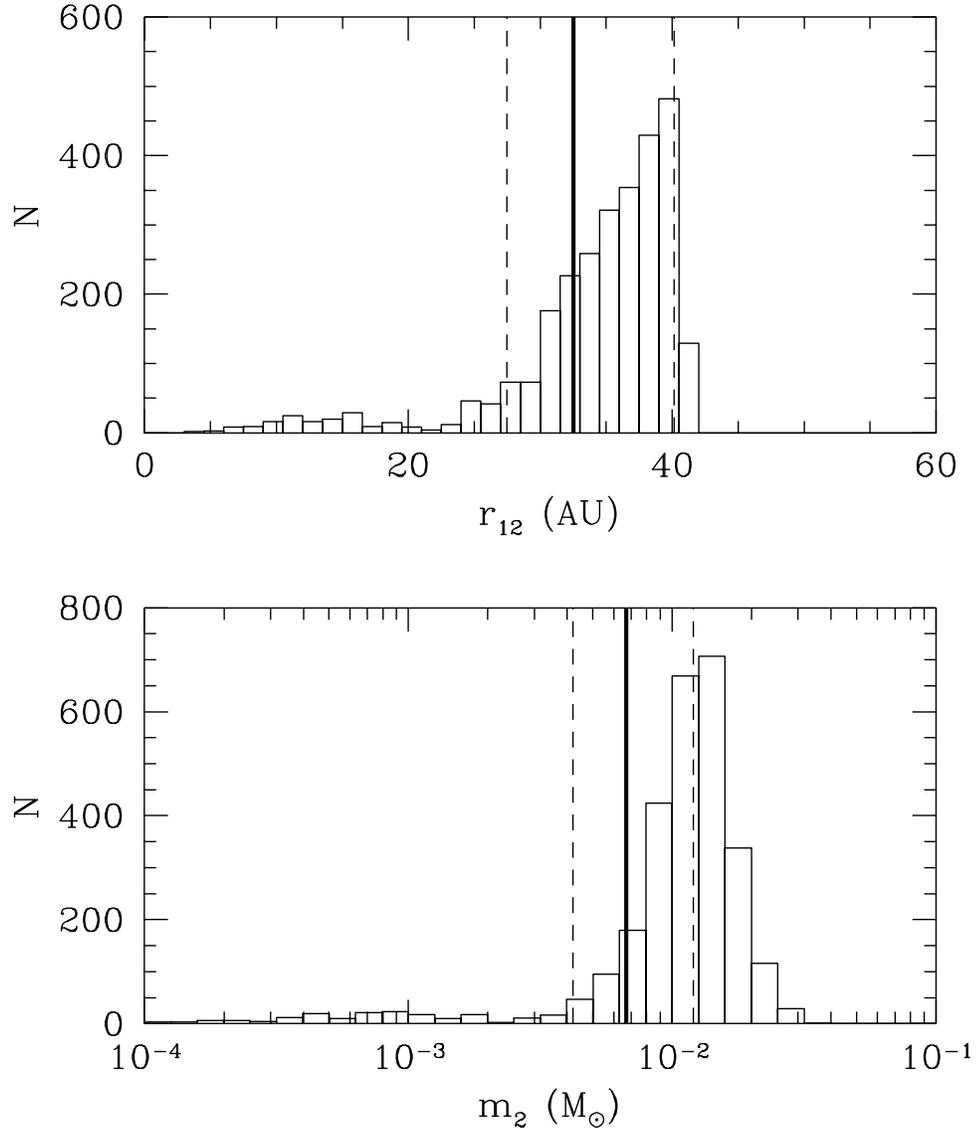}
\caption{Number of accepted realizations ($N$) of the triple for 
different values of $m_2$ and the corresponding distance $r_{12}$
of the second companion from the inner binary in our Monte-Carlo 
simulations. Accepted realizations are those leading to short-term
orbital perturbation effects consistent with the current observations.
The Monte-Carlo trials were performed using only
the standard one-parameter family of solutions obtained from the
first four pulse frequency derivatives. The solid line indicates the
value given by the preliminary measurement of $\fddddd$ and assuming
$\sin i_2=1$. 
The two dashed lines indicate the values obtained by decreasing 
the value of $\fddddd$ by a factor of 1.5 (right), or increasing 
it by a factor of 1.5 (left). 
\label{fig2}}
\end{figure}

\clearpage
\begin{figure}[t]
\plotone{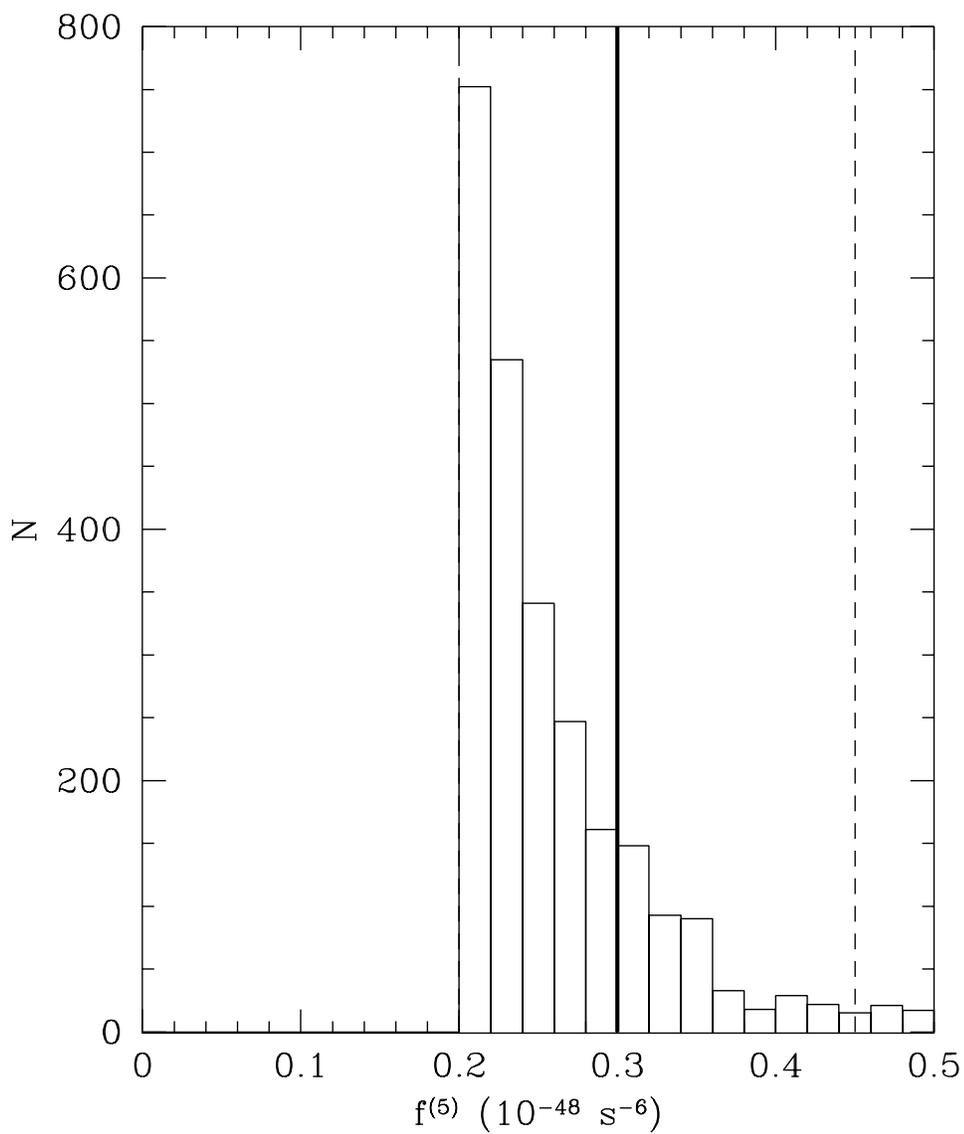}
\caption{Same as Fig.~2 but for the predicted value of the fifth
pulse derivative $\fddddd$. It is clear that the measured value
of $\fddddd$ is in perfect agreement with the theoretical expectations based
on the first four derivatives and the preliminary measurements of
orbital perturbation effects.
\label{fig3}}
\end{figure}

\clearpage
\begin{figure}[t]
\plotone{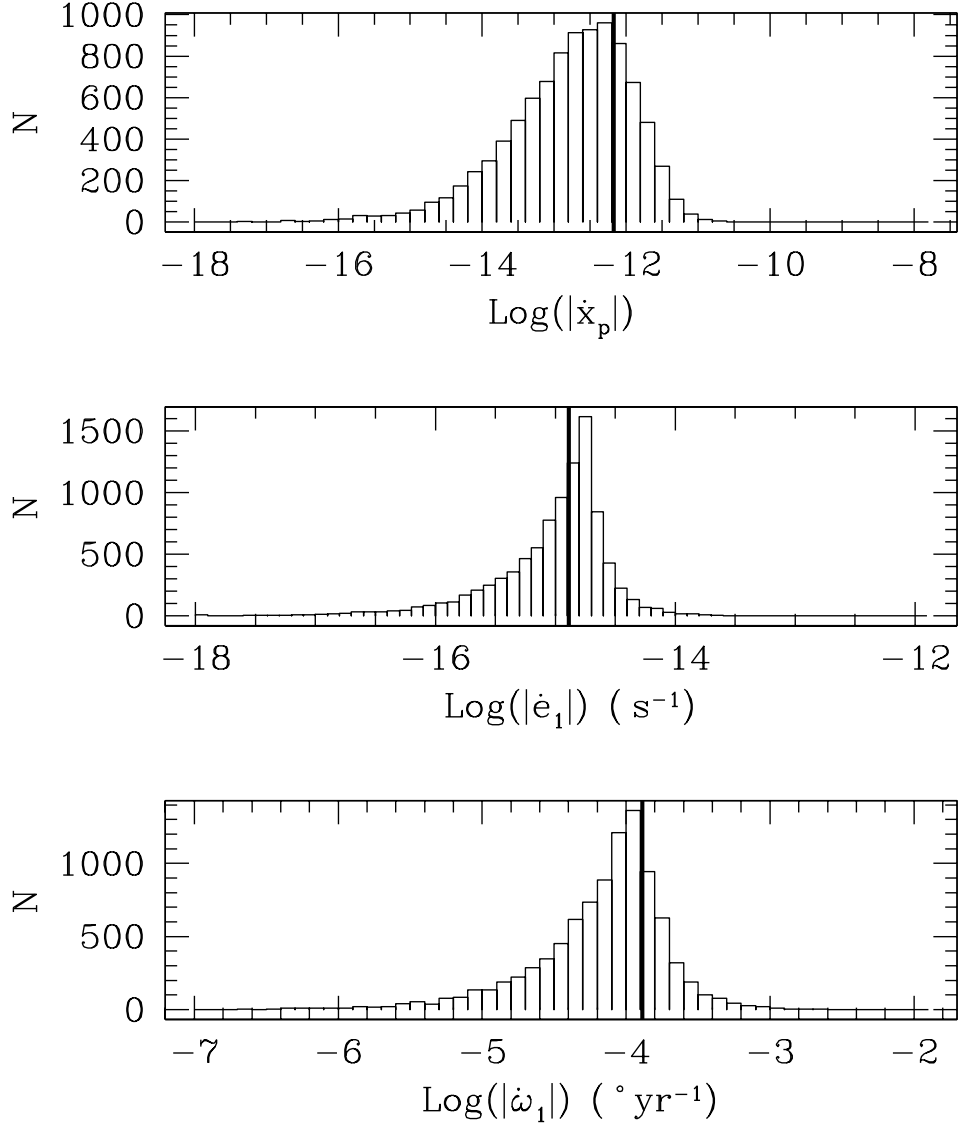}
\caption{A priori probability distributions of the three secular perturbations, 
eqs.~(2)--(4), based on the one-parameter family of solutions shown in Fig.~1 
and using \emph{only\/} the preliminary measurement of $\fddddd$
as a constraint in the Monte-Carlo simulations. 
The vertical lines show the measured value (for $\dot x_p$), or the 
$1\,\sigma$ upper limits on the measured values (for $\dot e_1$ and 
$\dot {\omega_1}$), which are all clearly consistent with these 
theoretical distributions.
\label{fig4}}
\end{figure}

\clearpage
\begin{figure}[t]
\plotone{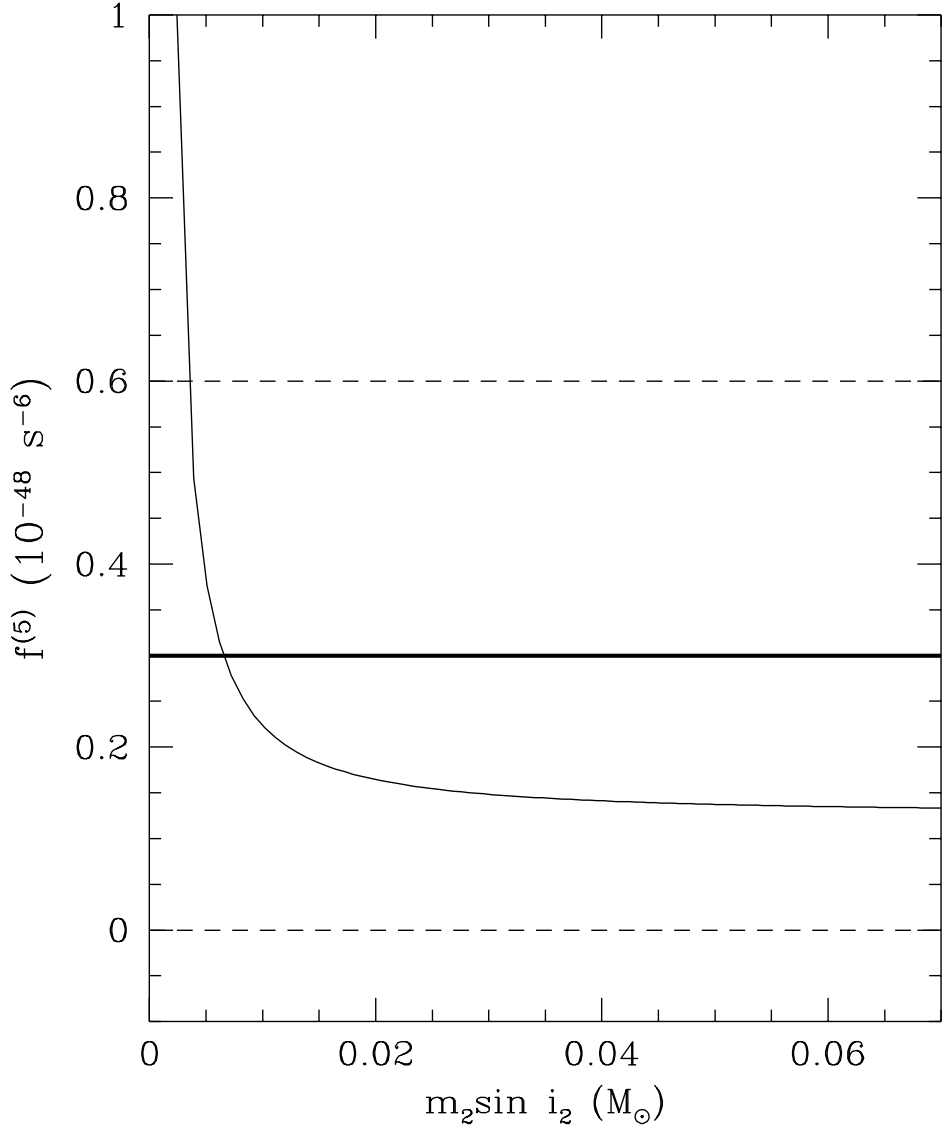}
\caption{Theoretically predicted value of $\fddddd$ as a function of the mass of
the second companion ($m_2$) for the one-parameter family of solutions
based on the first four derivatives. Here we have extended these solutions
into the hyperbolic regime. Values of $m_2\sin i_2 > 0.012\,M_\odot$ correspond 
to a hyperbolic orbit for the second companion ($e_2 > 1$). The solid 
horizontal line indicates the measured value of $\fddddd$, and the dashed
lines correspond to the current $1\,\sigma$ error. Thus the current
measurement uncertainty allows all hyperbolic solutions for the
second companion, and hence does not directly constrain $m_2$. However,
for $m_2 > 0.055\,M_\odot$, the relative velocity (``at infinity'')
of the hyperbolic orbit would exceed the escape
speed from the cluster core ($\simeq 12\,\rm{km\,s^{-1}}$).
\label{fig5}}
\end{figure}

\clearpage
\begin{figure}[t]
\plotone{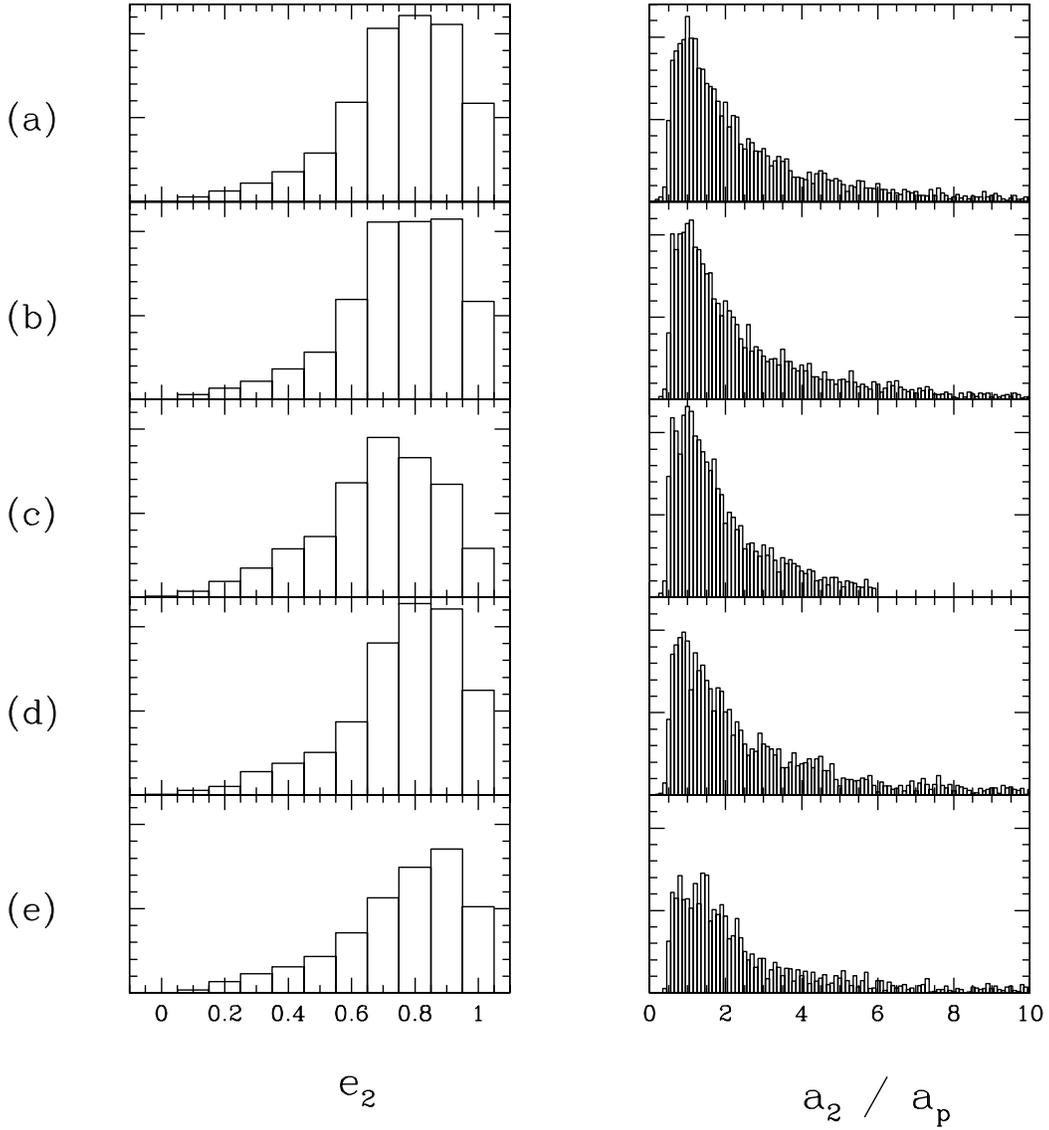}
\caption{Distributions of the final eccentricity $e_2$ and semimajor axis $a_2$ 
following an exchange interaction in which a planet of mass $m_p$
is captured by the binary pulsar. The semimajor axis in given in units of the original
star--planet separation $a_p$. Values of $m_p$ and $a_p$ are 
(a) $0.001\,M_\odot$, $5\,$AU; (b) $0.01\,M_\odot$, 
$5\,$AU; (c) $0.1\,M_\odot$, $5\,$AU; (d) $0.001\,M_\odot$, $30\,$AU;
(e) $0.01\,M_\odot$, $50\,$AU.
\label{fig6}}
\end{figure}

\clearpage
\begin{figure}[t]
\plotone{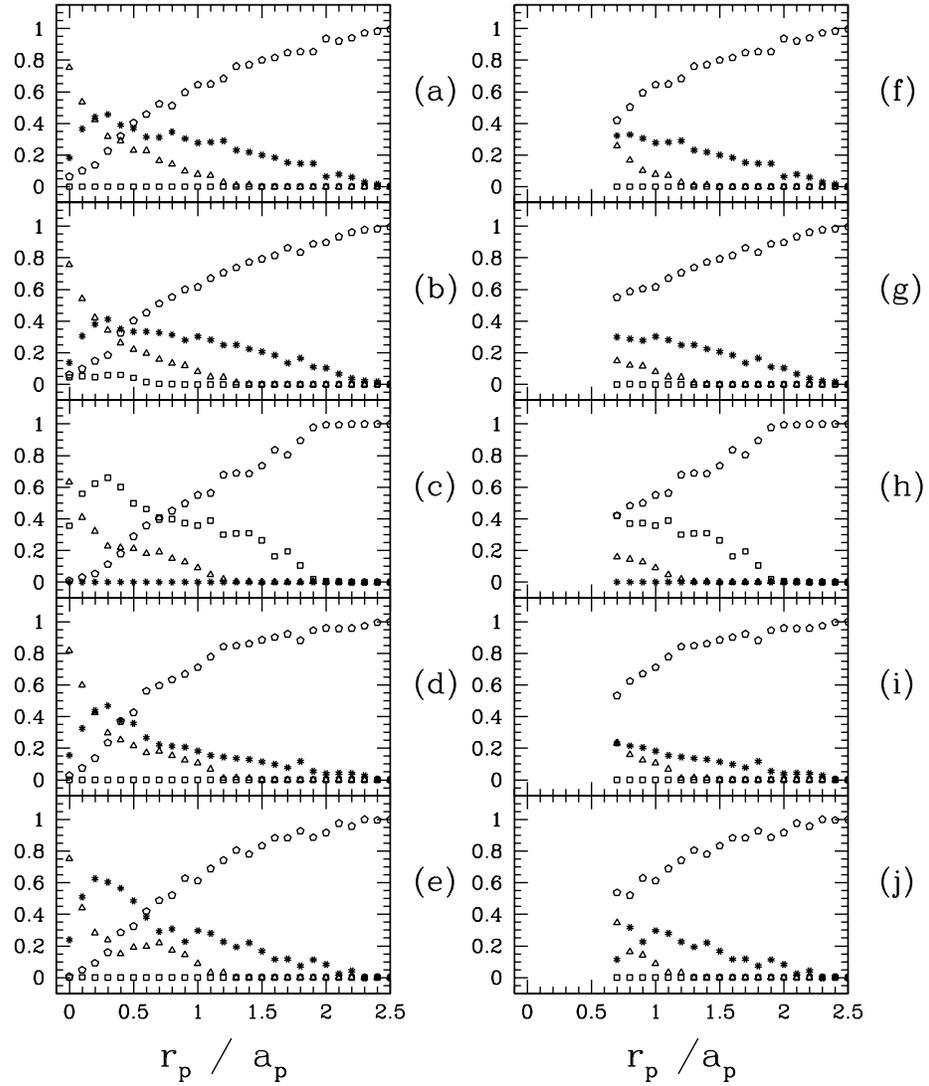}
\caption{Branching ratios for various possible outcomes
as a function of the distance of closest approach $r_p$ between 
the binary pulsar and the main-sequence star in the parent star--planet system.
The triangles represent the ejection of the star,
the squares the ejection of the planet, and the pentagons
a simple flyby with no exchange. Here the values of $m_p$ and
$a_p$ are: (a,f) $0.001\,M_\odot$, $5\,$AU; (b,g) $0.01\,M_\odot$, 
$5\,$AU; (c,h) $0.1\,M_\odot$, $5\,$AU; (d,i) $0.001\,M_\odot$, $30\,$AU; 
(e,j) $0.01\,M_\odot$, $50\,$AU. 
In panels (a)--(e) the only constraint imposed is that the binary
pulsar was not disrupted during the encounter (as in Fig.~6); 
in panels (f)--(j) a
stronger constraint was imposed, namely that the eccentricity of the
binary pulsar was not perturbed to a value exceeding the presently
observed value of 0.025. In all cases, we note that the branching ratio
for capture of the planet remains significant, at about 10--30\%. 
\label{fig7}}
\end{figure}

\clearpage
\begin{figure}[t]
\plotone{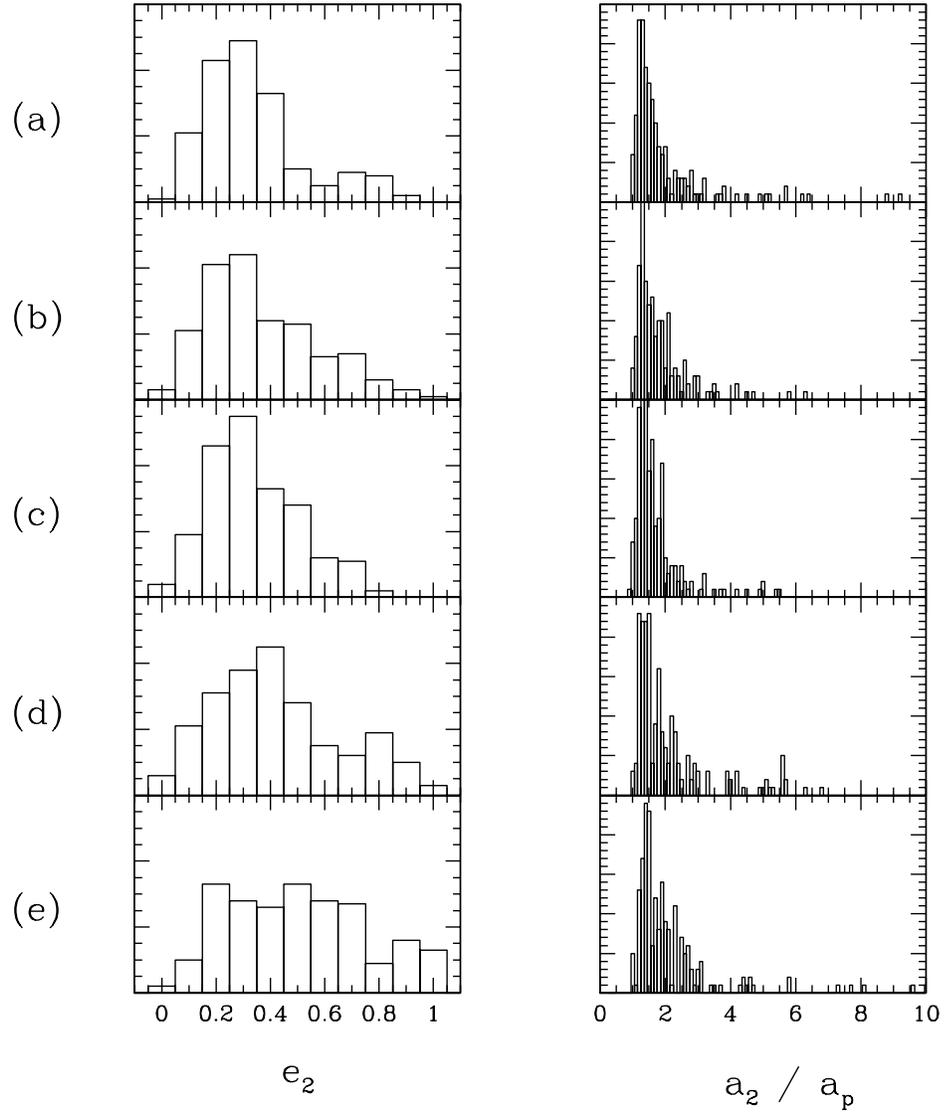}
\caption{Same as Fig.~6 but with the additional constraint that
the eccentricity of the binary pulsar was not perturbed to a value 
exceeding the presently observed value of 0.025.
\label{fig8}}
\end{figure}

\clearpage
\begin{figure}[t]
\plotone{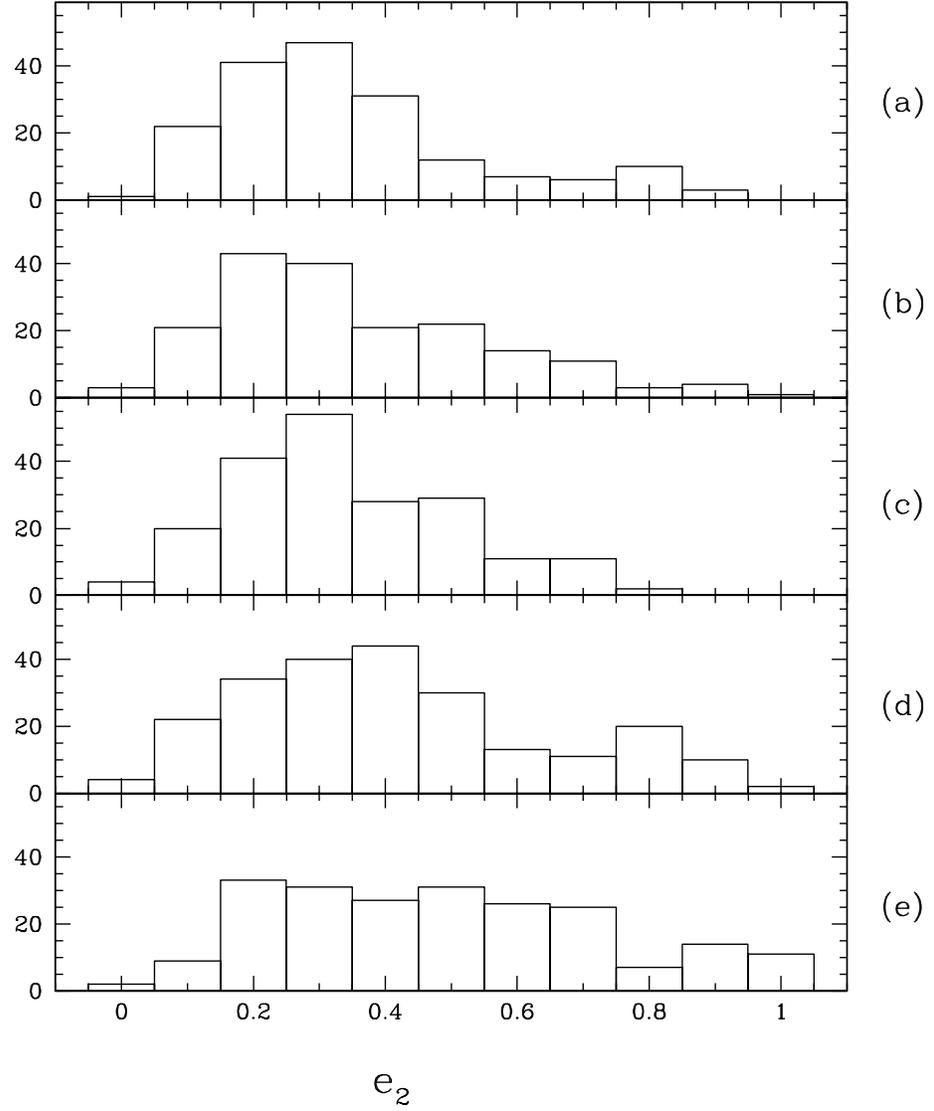}
\caption{Distribution of the eccentricity induced in the binary
pulsar during a dynamical exchange interaction in which a planet
was captured. The panels are labeled as in Fig.~6.
\label{fig9}}
\end{figure}

\clearpage
\begin{figure}[t]
\plotone{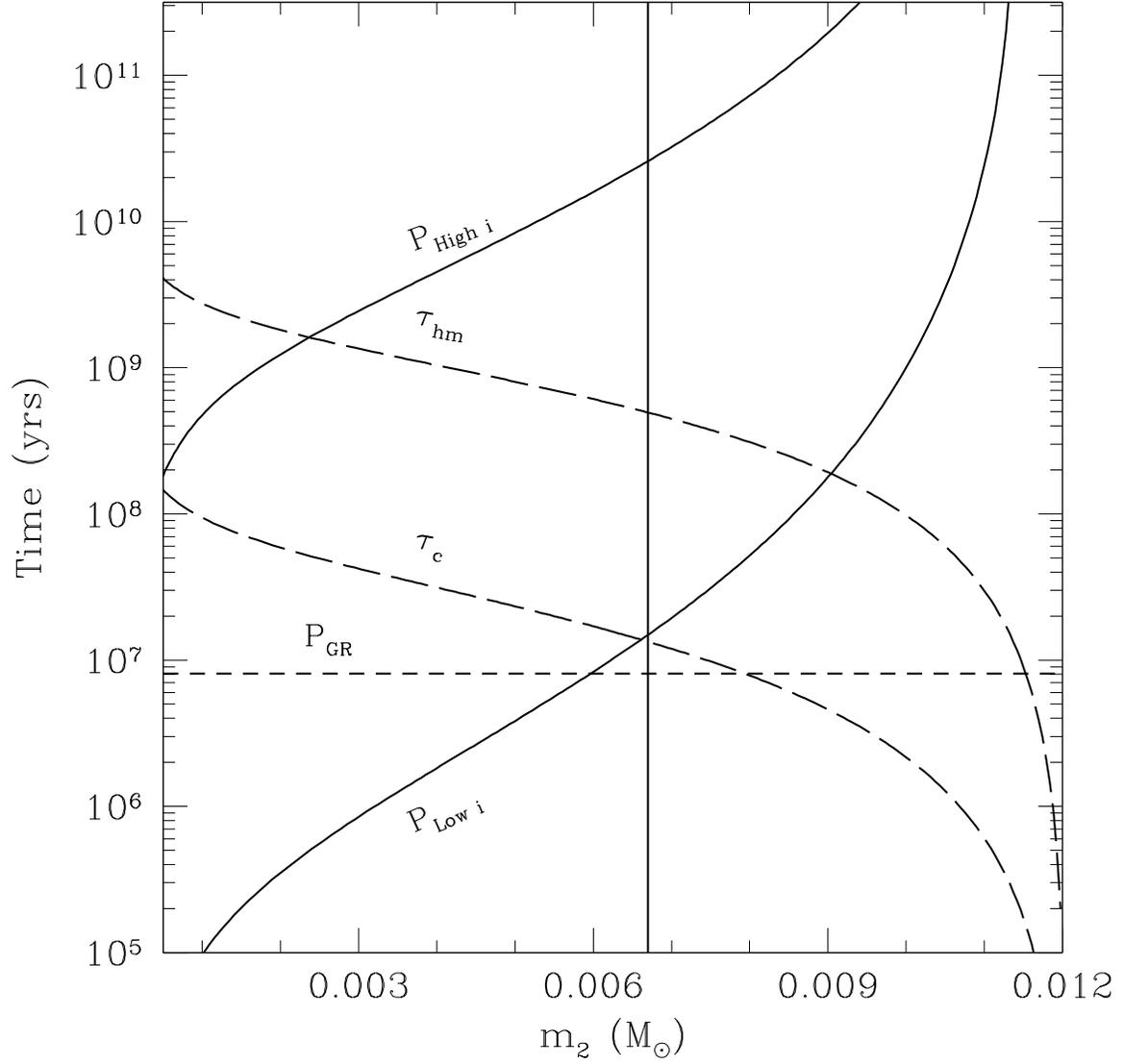}
\caption{Comparison of the various secular precession timescales 
in the PSR~B1620$-$26 triple, as a
function of the mass of the second companion in the standard solution of
\S 2. See text for details.
\label{fig10}}
\end{figure}

\clearpage
\begin{figure}[t]
\plotone{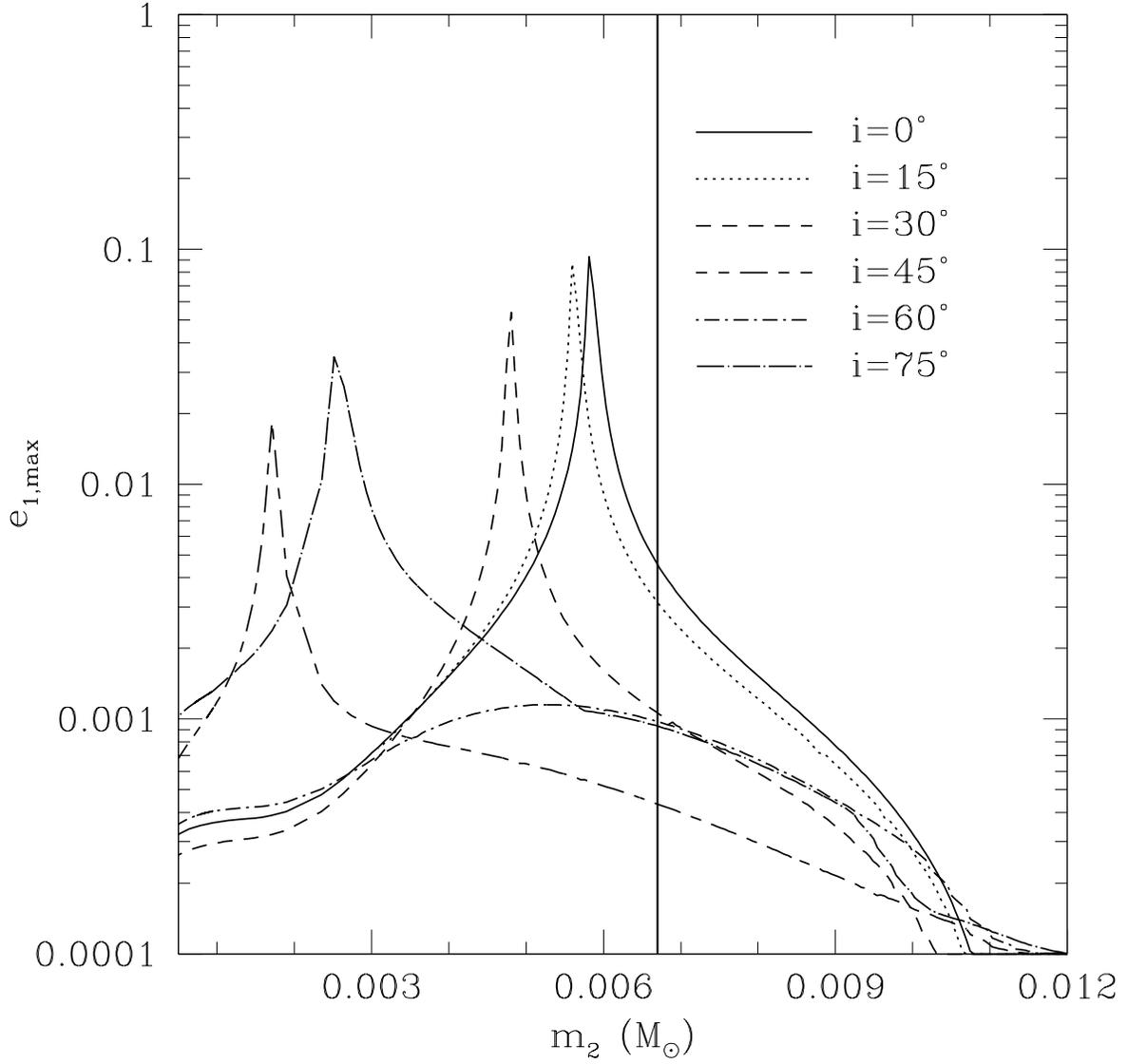}
\caption{Maximum eccentricity of the binary pulsar induced by secular
perturbations in the triple, as a
function of the mass of the second companion in the standard solution of
\S 2. See text for details.
\label{fig11}}
\end{figure}

\end{document}